\documentclass[journal]{IEEEtran}

\usepackage{amsmath,amssymb,amsfonts}
\usepackage{bm}
\usepackage{dsfont}
\usepackage{mathrsfs}
\usepackage{mathdots}
\usepackage{yhmath}
\usepackage{cancel}
\usepackage{braket}

\usepackage{graphicx}
\usepackage{tikz}
\usepackage{pgfplots}
\pgfplotsset{compat=1.16}

\usetikzlibrary{
    arrows.meta,
    calc,
    fadings,
    patterns,
    shadows.blur
}

\usepackage[
    font=footnotesize,
    labelfont=normalfont,
    textfont=normalfont
]{subcaption}

\usepackage{algorithm}
\usepackage{algpseudocode}

\usepackage{array}
\usepackage{tabularx}
\usepackage{booktabs}
\usepackage{multirow}
\usepackage{siunitx}

\newcolumntype{Y}{>{\raggedright\arraybackslash}X}

\usepackage{cite}
\usepackage{balance}
\usepackage{stfloats}
\usepackage{enumitem}
\usepackage{textcomp}
\usepackage{gensymb}
\usepackage[normalem]{ulem}

\usepackage{xcolor}

\usepackage[hidelinks]{hyperref}

\newtheorem{proposition}{Proposition}
\newtheorem{corollary}{Corollary}

\DeclareMathOperator{\erfc}{erfc}

\newcommand{\e}{\mathrm{e}}
\newcommand{\CRB}{\operatorname{CRB}}
\newcommand{\E}{\mathbb{E}}

\usepackage{amsmath,amssymb}
\usepackage[T1]{fontenc}
\usetikzlibrary{arrows.meta,calc,positioning,shapes.geometric}

\definecolor{navy}{HTML}{0B2E59}
\definecolor{teal}{HTML}{006D70}
\definecolor{orange}{HTML}{D95F02}
\definecolor{cloud}{HTML}{9DB7DF}
\definecolor{scatterteal}{HTML}{78B7B5}
\definecolor{scatterorange}{HTML}{F0A06B}
\definecolor{ink}{HTML}{151515}

\tikzset{
  >=Latex,
  panel/.style={draw=navy, line width=0.62pt, rounded corners=2.2pt,
                fill=white, inner sep=0pt},
  tealpanel/.style={draw=teal, line width=0.62pt, rounded corners=2.2pt,
                    fill=white, inner sep=0pt},
  latent/.style={draw=teal, line width=0.62pt, rounded corners=2.2pt,
                 dashed, dash pattern=on 2.2pt off 1.5pt,
                 fill=white, inner sep=0pt},
  flow/.style={draw=ink, line width=0.66pt,
               -{Latex[length=1.75mm,width=1.2mm]}},
  latentflow/.style={draw=teal, line width=0.66pt, dashed,
                     dash pattern=on 2.2pt off 1.5pt,
                     -{Latex[length=1.75mm,width=1.2mm]}},
  axis/.style={draw=ink, line width=0.40pt,
               -{Latex[length=1.35mm,width=0.95mm]}},
  title/.style={font=\sffamily\bfseries\scriptsize, text=navy,
                align=center},
  tealtitle/.style={font=\sffamily\bfseries\scriptsize, text=teal,
                    align=center},
  mathlabel/.style={font=\scriptsize, text=ink},
  minilabel/.style={font=\tiny, text=ink, align=center},
  signalzero/.style={fill=teal, draw=white, line width=0.32pt},
  signalone/.style={fill=orange, draw=white, line width=0.32pt}
}

\begin{document}
%\IEEEoverridecommandlockouts

\title{Quantum Communication and Phase Sensing \\over a Dephasing Channel with Thermal Noise}
% SEP and CRB Analysis of BPSK with Heterodyne Receiver over Dephasing Quantum Channels
\author{Nikos A. Mitsiou, Member, IEEE, and Ioannis Krikidis, Fellow, IEEE 

\thanks{Nikos A. Mitsiou and Ioannis Krikidis are with the Department of Electrical and Computer Engineering, University of Cyprus, 1678 Nicosia, Cyprus (e-mail: nmitsi02@ucy.ac.cy, krikidis@ucy.ac.cy).}

\vspace{-0.4cm}}

\maketitle
\begin{abstract}
Dephasing is a prominent source of decoherence in quantum communication, degrading phase coherence and transmission reliability. Therefore, knowledge of the physical channel parameters for reliable transmission is imperative, motivating the integration of quantum communication and sensing at the physical layer. Consequently, we study the transmission of coherent states over a von Mises dephasing channel affected by thermal noise. We consider a heterodyne receiver, for which, we first derive the symbol-error probability (SEP) and an accurate closed-form approximation, through which an irreducible asymptotic error floor due to the dephasing channel is identified. We note that the derived SEP is affected by the mismatch between the receiver's mean phase estimation and the true mean phase, thereby formalizing the interplay between sensing accuracy and communication reliability. In addition, we derive the Fisher information (FI) for the mean phase acquisition of the channel and provide closed-form lower and upper FI bounds. Moreover, the quantum FI (QFI) of the channel is studied, and lower and upper bounds on the QFI are given, which become asymptotically tight as the photon number increases. Despite the initial complexity of the FI and the QFI, the bounds given are relatively simple and effective. In addition, by comparing the FI with the QFI, we show that heterodyne detection approaches the quantum sensing limit as the number of photons increase, while an unavoidable sensing floor exists due to dephasing. Then, using the SEP and the FI analysis, we discuss the conditions under which communication and phase estimation requirements are jointly feasible. Finally, numerical results validate the derived SEP, estimation bounds, and the sensing-communication feasibility tradeoff.
\end{abstract}

\section{Introduction}
Integrated sensing and communication (ISAC) is regarded as a key enabler of future wireless systems, in which communication signals are designed not only to convey information but also to sense and interact with the surrounding environment \cite{zhang2021overview}. Accordingly, ISAC has been studied from several perspectives, including information theory, physical-layer design, networking, and practical implementation \cite{liu2022integrated}. At the heart of these efforts lies an inherent tradeoff, that of communication and sensing placing different demands on the transmitted signal and the available system resources, so improving one function may come at the expense of the other.

Meanwhile, quantum communication has emerged as another promising enabler of future networks, offering capabilities beyond those of conventional systems by exploiting quantum mechanical principles such as superposition, entanglement, quantum measurement and encoding classical information into quantum states of light \cite{djordjevic2022quantum}.  Although much of the interest in this field has been driven by quantum key distribution, quantum optical links are also being investigated from a broader communication-theoretic perspective. In fact, quantum optical states and measurements have been shown to improve the energy efficiency of optical communication protocols by enabling discrimination of optical coherent states with the discrimination error rate below the shot-noise limit \cite{burenkov2021practical}. In addition quantum states have been shown to increase the sensitivity of parameter estimation, supporting emerging quantum sensing applications, such as the quantum radar \cite{hanzo2025quantum}. 
% For example, the work in \cite{jung2025applying} applies turbo codingto coherent-state transmission over multipath channels affected by intersymbol interference, whereas \cite{krikidis2025tesla} develops a wireless-powered, batteryless quantum optical link and studies its fundamental detection and rate performance. 

Despite its potential, the performance of quantum communication is heavily affected by quantum channel uncertainty and decoherence, particularly when information is encoded in the optical phase \cite{mele2024quantum}. Thus, knowledge of the physical channel parameters for reliable transmission is crucial, which motivates the investigation of quantum integrated sensing and communication (QISAC), in which quantum resources are used to support both communication and sensing \cite{giani2025quantum}. Whereas classical ISAC balances communication rate, power consumption, and sensing accuracy under classical signal and noise models, QISAC must additionally account for quantum-state preparation, quantum noise, and the probabilistic nature of quantum measurements. Consequently, QISAC exhibits communication and sensing trade-offs akin to those encountered in ISAC, but accompanied by constraints and potential advantages that stem from the use of quantum states, measurements and entanglement \cite{monemi2026quantum}.

\subsection{Literature review}
Quantum communication with simultaneous channel parameter acquisition has recently attracted attention, particularly, from an information-theoretic perspective. In \cite{pereg2022communication}, capacity-distortion regions were derived for random-parameter quantum channels when the receiver must also reconstruct the channel state, while \cite{wang2022joint} characterized the tradeoff between reliable communication and parameter detection accuracy over classical-quantum channels. This framework was extended in \cite{munar2024joint}, where communication and reflectivity detection were studied over a lossy bosonic channel, and in \cite{erdogan2025joint}, where bipartite entanglement was allocated between communication and sensing over a thermal-loss bosonic channel. Recently, also, \cite{gong2025quantum} investigated joint optical communication and detection of transmissivity, showing how entanglement can reduce sensing latency while preserving communication performance. In addition, QISAC has also been investigated through the lens of explicit receiver utilization at the detection side. In \cite{liu2024quantum}, entanglement transmission was used to combine secure communication with phase sensing, while \cite{nikoloska2025variational} employed non-adaptive variational quantum measurements to balance communication throughput and parameter estimation accuracy. Moreover, \cite{krikidis2026quantum} considered the joint symbol detection and phase estimation of coherent states with homodyne receiver over a thermal Gaussian channel with unknown and deterministic phase rotation. Furthermore, phase estimation limits for displaced thermal states and practical Gaussian measurements were studied in \cite{aspachs2009phase}, whereas the simultaneous estimation of phase and phase-diffusion strength was considered in \cite{vidrighin2014joint}. Moreover,  phase estimation for a Gaussian channel with deterministic phase diffusion, but without thermal noise, was studied in \cite{genoni2011optical}, where homodyne detection was shown to approach the ultimate precision in specific noise regimes. Finally, \cite{huang2024exact} derived ultimate discrimination and estimation limits for  bosonic dephasing channels under ideal quantum receivers, nonetheless, the interplay between a practical communication and sensing receiver was not investigated.

\subsection{Motivation \& Contribution}
In optical communication links, atmospheric turbulence can produce wavefront distortions that phase compensation techniques partially suppress, leaving a residual phase error that can degrade communication performance \cite{zuo2011performance}. Furthermore, dephasing channels also describe another situation often encountered in satellite communication that of the sender and the receiver do not sharing a common phase reference during the transmission stage \cite{fanizza2021squeezing}.  All these motivate the study of coherent states transmission  over  dephasing channels in which communication and phase estimation are bound to the same resources. Despite that, the interplay between phase sensing and quantum communication is not sufficiently characterized. For instance, \cite{krikidis2026quantum}, which is the QISAC setting conceptually closest to ours, considered only an unknown deterministic phase rotation. In contrary, \cite{huang2024exact} studied stochastic phase diffusion channels and provided optimal estimation bounds, but it assumed ideal quantum receivers, and did not explore the communication-sensing tradeoff, neither thermal noise. Still, thermal noise can be one of the usual limiting factors in quantum and deep-space optical communications \cite{erkmen2009}. 

Motivated by this gap,  we investigate the joint communication
and mean-phase acquisition using coherent states over a quantum channel with thermal noise and
random von Mises dephasing. At the receiver's side, a heterodyne measurement is employed which has to support both functions; observations corresponding to known
acquisition symbols are used to estimate the mean dephasing angle, while a hyperplane detector recovers
the  data, subject to any errors due to the imperfect mean-phase estimation. To this end,  we
develop an analytical framework that characterizes the symbol error probability (SEP) of the communication stage and the Fisher information (FI) of the sensing stage. We also establish the sensing limits for this channel by deriving the quantum Fisher information (QFI). This work
further identifies the performance floors of communication and sensing that are caused by the random dephasing
and determines the conditions under which two communication and sensing
requirements can be simultaneously satisfied under a common photon allocation
budget. The main contributions are summarized as follows:

\begin{itemize}

    \item Since the exact SEP of the heterodyne
     detector under dephasing and thermal noise cannot be obtained in closed form, we provide a two point Gauss-Legendre
    approximation. From this analysis an error floor in the high photon regime is also identified that stems from dephasing. 

    \item For the mean phase estimation, we calculate the FI, and a lower and upper bound are provided. Despite the initial complexity of the FI, the bounds provided are simple, facilitating further analysis. It is also shown that as the number of photons increases, the heterodyne receiver reaches a sensing floor imposed by the dephasing channel's concentration around the mean. 

    % \item To fully characterize the ability of extracting information from the investigated channel we evaluate its QFI. To this end, we construct the Fock-basis representation of the phase shifted displaced thermal state and numerically evaluate its QFI. Rigorous bounds are derived, and by the QFI and FI comparison we deduce that the heterodyne receiver catches the QFI as the number of photons increases.
    \item Moreover, to characterize the ultimate information limit about the mean phase, independently of the receiver measurement, we evaluate the QFI. We also derive rigorous lower and upper QFI bounds and, by comparing the QFI with the heterodyne FI, we show that heterodyne detection asymptotically attains the quantum limit.

    \item For a sequential phase acquisition and communication protocol, we derive closed-form lower and upper photon allocation limits imposed by the sensing and SEP requirements. Together, these limits characterize the conditions for which both requirements are jointly feasible.

    \item Numerical results validate the SEP, the FI and QFI analysis, and the derived feasibility analysis for joint mean-phase sensing and communication.
\end{itemize}

\section*{Structure of the paper}
The structure of the paper is as follows. Section II describes the investigated channel model. Section III gives the SEP analysis of the heterodyne receiver. Section IV provides the CRB analysis of the heterodyne receiver. Section V presents the QFI analysis of the studied quantum channel, while Section VI discusses the tradeoff analysis between communication and sensing. Finally, Section VII numerically evaluates all the mathematical analysis and claims of the paper, while Section VIII gives conclusions and future research directions. 

\section{System model}
We consider a quantum optical link in which the transmitter
encodes classical information into coherent states using binary phase-shift keying (BPSK)\footnote{We note that our SEP analysis can be naturally extended to any constellation and modulation order. In addition, the sensing analysis of this paper is agnostic to the constellation used.}. Specifically, for each channel use, the transmitter prepares
a coherent state
\begin{equation}
    \ket{\alpha_s}
    =
    \ket{s\sqrt{N_s}},
    \qquad
    s\in\{-1,+1\},
    \label{eq:bpsk_states}
\end{equation}
where $N_s$ denotes the average photon number per symbol,
and the two symbols are equiprobable. We model the optical link as a 
channel with transmissivity $\eta\in(0,1]$ and thermal noise. In addition, the received field
is affected by a random phase rotation $\Phi$, which captures
dephasing due to path fluctuations, optical phase noise, partial phase mitigation, or communication without a common phase reference \cite{zuo2011performance, fanizza2021squeezing}.  To represent this uncertainty on the circular phase domain, we model the residual phase by a von Mises distribution \cite{vonmises}. Then, let $\Phi$ to follow a von Mises distribution with unknown mean phase $\Delta$ and known
concentration parameter $\kappa$, i.e.,
\begin{equation}
    p_{\Phi}(\phi;\Delta,\kappa)
    =
    \frac{\exp\!\left[\kappa\cos(\phi-\Delta)\right]}
    {2\pi I_0(\kappa)},
    \qquad
    -\pi\leq\phi<\pi,
    \label{eq:von_mises_phase}
\end{equation}
where $I_0(\cdot)$ denotes the modified Bessel function of
the first kind and order zero. We also note that the case of $\kappa\rightarrow\infty$ recovers the deterministic-phase model with $\Phi=\Delta$.

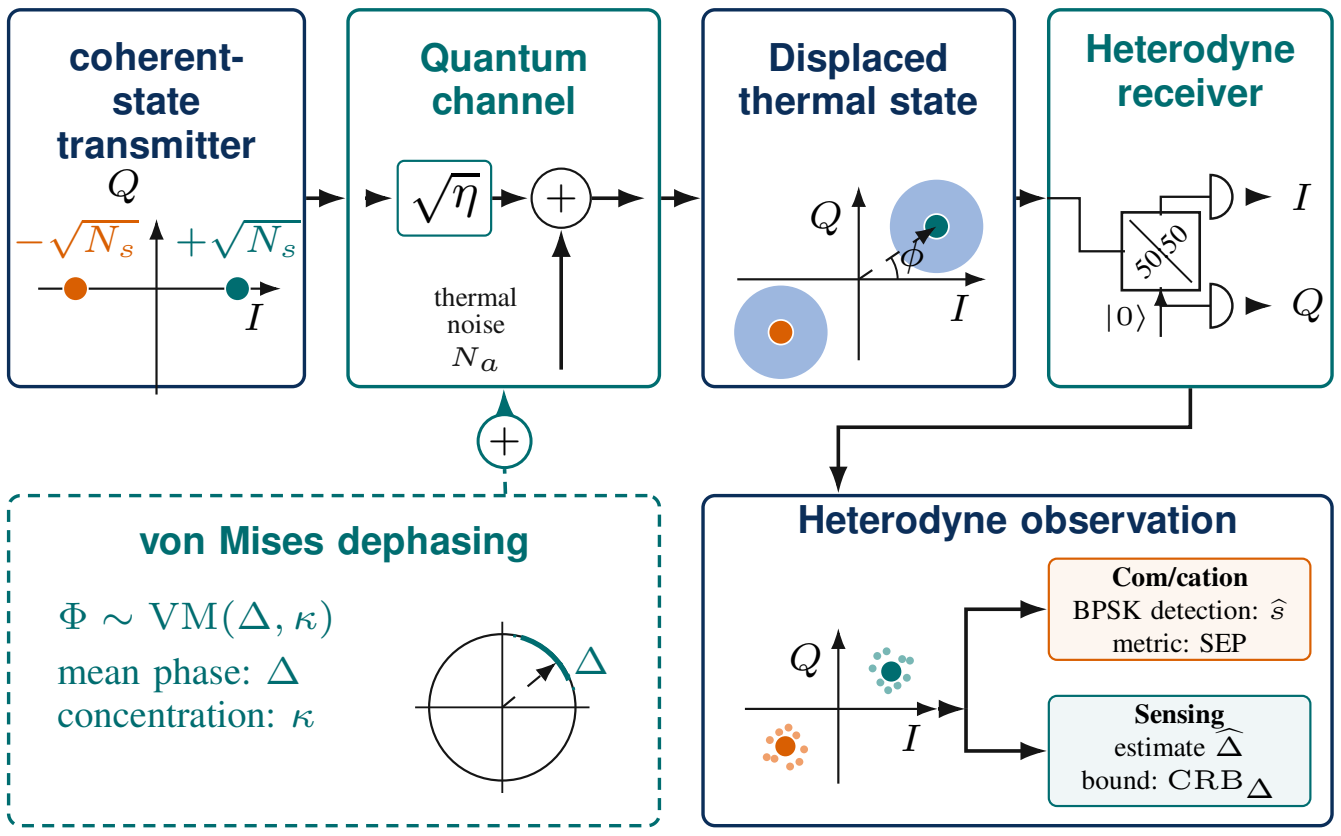
\begin{figure}[!t]
\centering
\resizebox{\columnwidth}{!}{%
\begin{tikzpicture}[x=1cm,y=1cm,font=\small]

\path[use as bounding box] (0,0) rectangle (8.60,5.40);

% =============================== TRANSMITTER ===============================
\node[
    panel,
    minimum width=1.90cm,
    minimum height=2.42cm,
    anchor=south west
] (tx) at (0.05,2.90) {};

\node[title,text width=1.68cm,anchor=north]
    at ($(tx.north)+(0,-0.10)$)
    {coherent-state\\transmitter};

\coordinate (tO) at ($(tx.south)+(0,0.64)$);

\draw[axis]
    ($(tO)+(-0.76,0)$) -- ($(tO)+(0.80,0)$)
    node[below left=-1pt,font=\scriptsize] {$I$};

\draw[axis]
    ($(tO)+(0,-0.68)$) -- ($(tO)+(0,0.45)$)
    node[above left=-1pt,font=\scriptsize] {$Q$};

\fill[signalone]
    ($(tO)+(-0.52,0)$) circle (2.25pt);

\fill[signalzero]
    ($(tO)+(0.52,0)$) circle (2.25pt);

\node[orange,above=2pt,font=\scriptsize]
    at ($(tO)+(-0.52,0)$)
    {$-\sqrt{N_s}$};

\node[teal,above=2pt,font=\scriptsize]
    at ($(tO)+(0.52,0)$)
    {$+\sqrt{N_s}$};

% ================================ CHANNEL =================================
\node[
    tealpanel,
    minimum width=2.00cm,
    minimum height=2.42cm,
    anchor=south west
] (ch) at (2.23,2.90) {};

\node[tealtitle,text width=1.78cm,anchor=north]
    at ($(ch.north)+(0,-0.10)$)
    {Quantum channel};

\draw[flow]
    ($(ch.west)+(0.12,0)$) --
    ($(ch.west)+(0.30,0)$);

\node[
    draw=teal,
    rounded corners=1.2pt,
    minimum width=0.60cm,
    minimum height=0.48cm,
    fill=white,
    inner sep=1pt
] (att) at ($(ch.west)+(0.63,0)$)
    {$\sqrt{\eta}$};

\draw[flow]
    (att.east) -- ($(ch.west)+(1.17,0)$);

\node[
    draw=ink,
    circle,
    minimum size=0.38cm,
    inner sep=0pt,
    fill=white,
    font=\scriptsize
] (plus) at ($(ch.west)+(1.38,0)$)
    {$+$};

\draw[flow]
    (plus.east) -- ($(ch.east)+(-0.10,0)$);

\draw[flow,-{Latex[length=1.8mm,width=1.25mm]}]
    ($(plus.south |- ch.south)+(0,0.12)$) --
    (plus.south);

\node[
    minilabel,
    font=\tiny,
    align=center,
    text width=0.76cm,
    anchor=south east,
    inner sep=0pt
] at ($(plus.south |- ch.south)+(-0.16,0.10)$)
    {thermal\\[-1pt]noise $N_a$};

% ========================= DISPLACED THERMAL STATE =========================
\node[
    panel,
    minimum width=2.00cm,
    minimum height=2.42cm,
    anchor=south west
] (state) at (4.51,2.90) {};

\node[title,text width=1.78cm,anchor=north]
    at ($(state.north)+(0,-0.10)$)
    {Displaced\\thermal state};

\coordinate (sO) at ($(state.south)+(0,0.70)$);

\draw[axis]
    ($(sO)+(-0.78,0)$) -- ($(sO)+(0.82,0)$)
    node[below left=-1pt,font=\scriptsize] {$I$};

\draw[axis]
    ($(sO)+(0,-0.52)$) -- ($(sO)+(0,0.60)$)
    node[below left=-1pt,font=\scriptsize] {$Q$};

\foreach \r/\op in {
    0.30/0.13,
    0.22/0.18,
    0.14/0.24,
    0.07/0.30
}{
    \fill[cloud,opacity=\op]
        ($(sO)+(0.50,0.34)$) circle (\r);

    \fill[cloud,opacity=\op]
        ($(sO)+(-0.50,-0.34)$) circle (\r);
}

\fill[signalzero]
    ($(sO)+(0.50,0.34)$) circle (2.25pt);

\fill[signalone]
    ($(sO)+(-0.50,-0.34)$) circle (2.25pt);

\draw[ink,dashed,-{Latex[length=1.35mm]}]
    (sO) -- ($(sO)+(0.50,0.34)$);

\draw[ink,line width=0.36pt]
    ($(sO)+(0.25,0)$)
    arc[start angle=0,end angle=34.2,radius=0.25];

\node[font=\scriptsize]
    at ($(sO)+(0.34,0.15)$)
    {$\phi$};

% ============================ HETERODYNE RECEIVER ==========================
\node[
    tealpanel,
    minimum width=1.82cm,
    minimum height=2.42cm,
    anchor=south west
] (rx) at (6.73,2.90) {};

\node[tealtitle,text width=1.60cm,anchor=north]
    at ($(rx.north)+(0,-0.05)$)
    {Heterodyne\\receiver};

\coordinate (bs) at ($(rx.south)+(-0.19,0.88)$);

\draw[ink,fill=white,line width=0.48pt]
    ($(bs)+(-0.25,-0.25)$)
    rectangle
    ($(bs)+(0.25,0.25)$);

\draw[ink,line width=0.36pt]
    ($(bs)+(-0.20,0.20)$) --
    ($(bs)+(0.20,-0.20)$);

\node[
    font=\tiny,
    rotate=45,
    inner sep=0pt
] at (bs)
    {50:50};

\draw[ink,line width=0.42pt,-{Latex[length=1.2mm]}]
    ($(bs)+(0,-0.55)$) --
    ($(bs)+(0,-0.23)$);

\node[font=\tiny,anchor=east,inner sep=0pt]
    at ($(bs)+(-0.06,-0.43)$)
    {$|0\rangle$};

\coordinate (iDet) at ($(rx.south)+(0.31,1.23)$);
\coordinate (qDet) at ($(rx.south)+(0.31,0.53)$);

\draw[ink,line width=0.42pt]
    (rx.west) -- ++(0.21,0)
    |- ($(bs)+(-0.23,0)$);

\draw[ink,line width=0.42pt]
    ($(bs)+(0,0.23)$)
    |- ($(iDet)+(-0.18,0)$);

\draw[ink,line width=0.42pt]
    ($(bs)+(0,-0.23)$)
    |- ($(qDet)+(-0.18,0)$);

\draw[ink,fill=white,line width=0.45pt]
    ($(iDet)+(-0.18,-0.14)$) --
    ($(iDet)+(-0.18,0.14)$)
    arc[start angle=90,end angle=-90,radius=0.14] --
    cycle;

\draw[ink,fill=white,line width=0.45pt]
    ($(qDet)+(-0.18,-0.14)$) --
    ($(qDet)+(-0.18,0.14)$)
    arc[start angle=90,end angle=-90,radius=0.14] --
    cycle;

\draw[flow]
    ($(iDet)+(0.10,0)$) -- ++(0.12,0);

\draw[flow]
    ($(qDet)+(0.10,0)$) -- ++(0.12,0);

\node[font=\scriptsize,anchor=west]
    at ($(iDet)+(0.21,0)$)
    {$I$};

\node[font=\scriptsize,anchor=west]
    at ($(qDet)+(0.21,0)$)
    {$Q$};

% ============================ DEPHASING MODEL ==============================
\node[
    latent,
    minimum width=4.18cm,
    minimum height=2.12cm,
    anchor=south west
] (phase) at (0.05,0.08) {};

\node[tealtitle,text width=3.80cm,anchor=north]
    at ($(phase.north)+(0,-0.08)$)
    {von Mises dephasing};

\node[text=teal,font=\scriptsize,align=left,anchor=west]
    at ($(phase.west)+(0.20,-0.02)$)
    {$\Phi\sim\mathrm{VM}(\Delta,\kappa)$\\[2pt]
     mean phase: $\Delta$\\
     concentration: $\kappa$};

\coordinate (pO) at ($(phase.east)+(-1.02,-0.30)$);

\draw[ink,line width=0.40pt]
    (pO) circle (0.47);

\draw[ink,line width=0.34pt]
    ($(pO)+(-0.56,0)$) -- ($(pO)+(0.56,0)$);

\draw[ink,line width=0.34pt]
    ($(pO)+(0,-0.55)$) -- ($(pO)+(0,0.55)$);

\draw[ink,dashed,-{Latex[length=1.35mm]}]
    (pO) -- ++(0.35,0.29);

\draw[teal,line width=0.95pt]
    ($(pO)+(0.12,0.45)$)
    arc[start angle=74,end angle=22,radius=0.47];

\draw[teal,dotted,line width=0.54pt]
    ($(pO)+(0.07,0.46)$)
    arc[start angle=82,end angle=12,radius=0.47];

\node[text=teal,font=\scriptsize]
    at ($(pO)+(0.57,0.30)$)
    {$\Delta$};

% ================== SHARED OBSERVATION AND QISAC TASKS =====================
\node[
    panel,
    minimum width=4.04cm,
    minimum height=2.12cm,
    anchor=south west
] (proc) at (4.51,0.08) {};

\node[title,text width=3.72cm,anchor=north]
    at ($(proc.north)+(0,0.05)$)
    {Heterodyne observation};

\coordinate (oO) at ($(proc.west)+(0.88,-0.31)$);

\draw[axis]
    ($(oO)+(-0.60,0)$) -- ($(oO)+(0.64,0)$)
    node[below left=-1pt,font=\scriptsize] {$I$};

\draw[axis]
    ($(oO)+(0,-0.48)$) -- ($(oO)+(0,0.52)$)
    node[below left=-1pt,font=\scriptsize] {$Q$};

\foreach \dx/\dy in {
    -0.12/0.02,
    -0.09/0.10,
    -0.07/-0.07,
    -0.03/0.13,
     0/0,
     0.04/-0.09,
     0.07/0.08,
     0.10/-0.04,
     0.12/0.07,
    -0.02/-0.12
}{
    \fill[scatterteal,opacity=0.68]
        ($(oO)+({0.34+\dx},{0.24+\dy})$)
        circle (0.72pt);

    \fill[scatterorange,opacity=0.68]
        ($(oO)+({-0.34-\dx},{-0.24-\dy})$)
        circle (0.72pt);
}

\fill[teal,draw=white,line width=0.32pt]
    ($(oO)+(0.34,0.24)$) circle (2.00pt);

\fill[orange,draw=white,line width=0.32pt]
    ($(oO)+(-0.34,-0.24)$) circle (2.00pt);

\node[
    draw=orange,
    fill=orange!6,
    rounded corners=1.5pt,
    minimum width=1.68cm,
    minimum height=0.62cm,
    text width=1.48cm,
    align=center,
    font=\tiny,
    inner sep=1.5pt
] (comm) at ($(proc.east)+(-0.99,0.33)$)
    {\textbf{Com/cation}\\
     BPSK detection: $\widehat{s}$\\
     metric: SEP};

\node[
    draw=teal,
    fill=teal!6,
    rounded corners=1.5pt,
    minimum width=1.68cm,
    minimum height=0.62cm,
    text width=1.48cm,
    align=center,
    font=\tiny,
    inner sep=1.5pt
] (sense) at ($(proc.east)+(-0.99,-0.58)$)
    {\textbf{Sensing}\\
     estimate $\widehat{\Delta}$\\
     bound: $\CRB_\Delta$};

\coordinate (fork)
    at ($(proc.west)+(1.70,-0.31)$);

\draw[flow]
    ($(oO)+(0.64,0)$) -- (fork);

\draw[flow]
    (fork) |- (comm.west);

\draw[flow]
    (fork) |- (sense.west);

% ============================== FLOW ARROWS ================================
\draw[flow]
    (tx.east) -- (ch.west);

\draw[flow]
    (ch.east) -- (state.west);

\draw[flow]
    (state.east) -- (rx.west);

\coordinate (phaseout)
    at ($(phase.north east)+(-1.00,0)$);

\node[
    draw=teal,
    circle,
    fill=white,
    line width=0.55pt,
    minimum size=0.34cm,
    inner sep=0pt,
    font=\scriptsize
] (phaseplus) at ($(phaseout)!0.5!(ch.south)$)
    {$+$};

\draw[
    teal,
    dashed,
    line width=0.66pt,
    dash pattern=on 2.2pt off 1.5pt
]
    (phaseout) -- (phaseplus.south);

\draw[latentflow]
    (phaseplus.north) -- (ch.south);

\draw[flow]
    (rx.south) -- ++(0,-0.25)
    -| ($(proc.north west)+(0.88,0)$);

\end{tikzpicture}%
}
\caption{The investigated QISAC model.}
\label{fig:system_model_bpsk_heterodyne}
\end{figure}

Conditioned on the phase realization $\Phi=\phi$, the channel
output corresponding to the BPSK input $\ket{\alpha_s}$ is a
displaced thermal state, which can be written as
\begin{equation}
    \rho_{\mathrm{out}}(s\mid\phi)
    =
    \frac{1}{\pi N_a}
    \int
    \exp\!\left(
        -\frac{\left\lvert\alpha-\beta_s(\phi)\right\rvert^2}
        {N_a}
    \right)
    \ket{\alpha}\bra{\alpha}\,d^2\alpha,
    \label{eq:conditional_thermal_state}
\end{equation}
%\mathbb{C}
where $
    \beta_s(\phi)
    =
    \sqrt{\eta}\alpha_s e^{\jmath\phi}
    =
    s\sqrt{\eta N_s}\,e^{\jmath\phi} $ is the mean field after attenuation and phase rotation, and $N_a$ is the mean photon number due to thermal noise. Accordingly, the output state after averaging over the dephasing process is
\begin{equation}
    \rho_{\mathrm{out}}(s)
    =
    \int_{-\pi}^{\pi}
    p_{\Phi}(\phi;\Delta,\kappa)\,
    \rho_{\mathrm{out}}(s\mid\phi)\,d\phi .
    \label{eq:qfi_received_state}
\end{equation}

At the receiver, we employ an ideal heterodyne detector, which jointly measures the in-phase and quadrature components of the received field. We note that, although heterodyne detection does not generally attain the ultimate quantum limit on optical communication, it is particularly well suited to thermal-noise channels. Specifically, in the high background noise and low signal levels, coherent homodyne and heterodyne receivers have been shown to approach the ultimate classical capacity of the single-mode thermal channel, while heterodyne detection remains near-optimal at high received signal levels \cite{erkmen2009}. Although these capacity results do not necessarily generalize to the von Mises dephasing model with thermal noise channel considered in this work, they provide a strong  motivation for adopting the heterodyne detection.

Then, conditioned on $s$ and $\Phi=\phi$, the complex
heterodyne output is modeled as
\begin{equation}
    Z
    =
    \beta_s(\phi)+W_{\mathrm{het}}
    =
    s\sqrt{\eta N_s}\,e^{\jmath\phi}
    +W_{\mathrm{het}},
    \label{eq:heterodyne_output}
\end{equation}
where $     W_{\mathrm{het}} \sim    \mathcal{CN}\!\left(0,N_a+1\right)$ is circularly symmetric complex Gaussian. The term $N_a$
accounts for the channel thermal noise, while the additional
unit accounts for noise due to the heterodyne measurement. For subsequent analysis, we define the real in-phase and
quadrature observation vector as~$\mathbf{Y}\!=\begin{bmatrix} \Re\{Z\}, \Im\{Z\} \end{bmatrix}^{\mathsf{T}} $. Hence, conditioned on $s$ and $\Phi=\phi$, the heterodyne output
follows the bivariate Gaussian distribution
$\mathbf{Y}
    \sim
    \mathcal{N}_2\!\left(
        \boldsymbol{\mu}_s(\phi),
        \sigma_{\mathrm{het}}^2\mathbf{I}_2
    \right)$,
where
    $\boldsymbol{\mu}_s(\phi)
    =
    s\sqrt{\eta N_s}\bm e_\phi,
    \,\,
    \sigma_{\mathrm{het}}^2
    =
    \frac{N_a+1}{2}$
and $\mathbf{I}_2$ is the $2\times2$ identity matrix. The vector $\bm e_\phi=[\cos\phi,\sin\phi]^{\mathsf T}$ denotes the unit
vector of direction $\phi$. Since the phase realization $\Phi$ is not directly observed at the
receiver, it acts as a latent random variable. Therefore, the probability density function (pdf) of the heterodyne observation, conditioned on symbol $s,\Delta, \kappa$, is
obtained by averaging the bivariate conditional Gaussian pdf over the von Mises
phase distribution in \eqref{eq:von_mises_phase}, i.e.,
\begin{equation}
    p(\mathbf{y}\mid s;\Delta,\kappa)
    =
    \int_{-\pi}^{\pi}
    p_{\Phi}(\phi;\Delta,\kappa)\,
    \mathcal{N}_2\!\left(
        \mathbf{y};
        \boldsymbol{\mu}_s(\phi),
        \sigma_{\mathrm{het}}^2\mathbf{I}_2
    \right)
    d\phi .
    \label{eq:heterodyne_marginal_likelihood}
\end{equation}
This marginal density characterizes the heterodyne observation
under each transmitted BPSK symbol and will be used to evaluate the communication and sensing capabilities of the heterodyne receiver in a von Mises dephasing channel with thermal noise.
% In the following section, for an observed vector $\mathbf{y}$, it is used as the likelihood function with respect to the unknown mean phase $\Delta$.

\section{SEP of the heterodyne detector}
In this section, we characterize the communication performance of
coherent state BPSK transmission. Following
\cite{krikidis2026quantum}, we consider a linear detector whose
decision boundary is a hyperplane through the origin. For the
heterodyne output \(\mathbf{Y}\), the detected symbol is
\begin{equation}
    \widehat{s}
    =
    \operatorname{sgn}\!\left(
        \bm e_\vartheta^{\mathsf T}\mathbf{Y}
    \right),
    \qquad
    \vartheta\in[-\pi,\pi).
    \label{eq:hyperplane_detector}
\end{equation}
Here, $\vartheta$ denotes the orientation of the detector, which can be viewed as the receiver's current prediction of the mean dephasing value $\Delta$. The optimal alignment of the detector is $\vartheta^*=\Delta$. However, since the $\Delta$ is not perfectly known, let us assume that $e=\Delta-\vartheta$ denotes the residual angular mismatch of the detector. Then, conditioned on $s,\Phi=\phi$, we define the scalar decision
statistic $T    = \bm e_\vartheta^{\mathsf T}\mathbf{Y}$ and we obtain
\begin{equation}
\begin{aligned}
    T
    &=
    s\sqrt{\eta N_s}
    \bm e_\vartheta^{\mathsf T}\bm e_\phi
    +
    \bm e_\vartheta^{\mathsf T}\mathbf{W}_{\mathrm{het}}  \\
    &=
    s\sqrt{\eta N_s}\cos(\phi-\vartheta)
    +
    \bm e_\vartheta^{\mathsf T}\mathbf{W}_{\mathrm{het}}.
\end{aligned}
\end{equation}
Since $\bm e_\vartheta$ is a unit vector, the projected noise $ \bm e_\vartheta^{\mathsf T}\mathbf{W}$ remains
Gaussian with the same variance $\sigma_{\mathrm{het}}^2$. Hence, conditioned on $s$ and $\Phi=\phi$, $T\sim
    \mathcal{N}\!\left(
        s\sqrt{\eta N_s}\cos(\phi-\vartheta),
        \sigma_{\mathrm{het}}^2
    \right)$.
 For $s=+1$, an error occurs when $T<0$. In contrast, for $s=-1$, an error occurs when $T\geq0$. Thus, due to symmetry, both BPSK symbols have the same conditional error
probability, which can be easily shown to be given as
\begin{equation}
\begin{aligned} 
    \Pr(\widehat{s}\neq s\mid\Phi=\phi)
    &=
    Q\!\left(
        \frac{
            \sqrt{\eta N_s}\cos(\phi-\vartheta)
        }{
            \sigma_{\mathrm{het}}
        }
    \right).
\end{aligned}
\label{eq:conditional_sep_positive_symbol}
\end{equation}
Thus, the detector's performance is directly affected by the estimation accuracy of $\Delta$. Averaging this
conditional error probability with respect to the von Mises
density in \eqref{eq:von_mises_phase}, the SEP of the heterodyne detection is given as 
\begin{equation}
    P_{\rm e}
    =
    \int_{-\pi}^{\pi}
    p_{\Phi}(\phi;\Delta,\kappa)
    Q\!\left(
        \frac{
            \sqrt{\eta N_s}\cos(\phi-\vartheta)
        }{
            \sigma_{\mathrm{het}}
        }
    \right)
    d\phi .
    \label{eq:heterodyne_hyperplane_sep}
\end{equation}
We note that $\eqref{eq:heterodyne_hyperplane_sep}$ cannot be obtained in closed form with respect to known elementary functions, thus, to proceed with our analysis, we propose to approximate the SEP using the widely used Gauss-Legendre (GL) approximation. The two-point GL approximation of the hyperplane detector's SEP is given by the following proposition. 

% \begin{proposition}
% The two-point Gauss-Legendre approximation of the hyperplane detector SEP is given as
% \begin{equation}
%     P_{\rm e}(\rho)
%     \approx
%     P_{\infty}
%     +
%     \Gamma_1Q(c_1\sqrt{\rho})
%     +
%     \Gamma_2Q(c_2\sqrt{\rho}),
%     \label{eq:sep_two_point_gl}
% \end{equation}
% where   $\rho = \frac{\eta N_s}{\sigma_{\mathrm{het}}^2}$, $e = \Delta-\vartheta$, $c_i=\cos x_i$,     $x_1  =   \frac{\pi}{4}\left(1-\frac{1}{\sqrt{3}}\right)$, $x_2 = \frac{\pi}{4}\left(1+\frac{1}{\sqrt{3}}\right)$, $ \Gamma_i
%     =
%     \frac{
%         \sinh(\kappa_{\rm c}\cos x_i)
%         \cosh(\kappa_{\rm s}\sin x_i)
%     }{2I_0(\kappa)},
%     \, i\in\{1,2\}.$
% and 
% \begin{equation}
%     P_{\infty}
%     =
%     \frac{1}{2\pi I_0(\kappa)}
%     \int_{-\frac{pi}{2}}^{\frac{\pi}{2}}
%     \exp\!\left(-\kappa\cos(x-e)\right)dx .
%     \label{eq:sep_phase_floor_positive_coordinate}
% \end{equation}
% \end{proposition}
\begin{proposition}
\label{prop:two_point_gl_sep}
Let $\rho=\eta N_s/\sigma_{\mathrm{het}}^2$, $e=\Delta-\vartheta$,
$\kappa_{\rm c}=\kappa\cos e$, and $\kappa_{\rm s}=\kappa\sin e$. For $\kappa>0$ and
$|e|<\pi/2$, the two-point GL approximation of the
hyperplane-detector SEP is given as
\begin{equation}
    P_{\rm e}(\rho)
    \approx
    P_{\infty}
    +
    \Gamma_1Q(c_1\sqrt{\rho})
    +
    \Gamma_2Q(c_2\sqrt{\rho}),
    \label{eq:sep_two_point_gl}
\end{equation}
where $x_1=\frac{\pi}{4}\left(1-\frac{1}{\sqrt{3}}\right)$,
$x_2=\frac{\pi}{4}\left(1+\frac{1}{\sqrt{3}}\right)$,
$c_i=\cos x_i$, and
\[
    \Gamma_i
    =
    \frac{
        \sinh(\kappa_{\rm c}\cos x_i)
        \cosh(\kappa_{\rm s}\sin x_i)
    }{
        2I_0(\kappa)
    },
    \qquad i\in\{1,2\},
\]
and the phase-error floor, $ P_{\infty}$, is defined as
\begin{equation}
    P_{\infty}
    =
    \frac{1}{2\pi I_0(\kappa)}
    \int_{-\pi/2}^{\pi/2}
    \exp\!\left[-\kappa\cos(x-e)\right]dx .
    \label{eq:sep_phase_floor_positive_coordinate}
\end{equation}
\end{proposition}

\begin{IEEEproof}
See Appendix~\ref{app:two_point_gl_sep_proof}.
\end{IEEEproof}
\begin{corollary}
To interpret $P_{\infty}$, let
$U=\Phi-\Delta\sim\mathrm{VM}(0,\kappa)$.  With exception at the boundary points satisfying $\cos(U+e)=0$, which have
zero probability under the continuous von Mises distribution,
\[
    Q\!\left(\sqrt{\rho}\cos(U+e)\right)
    \longrightarrow
    \mathds{1}\{\cos(U+e)<0\}
    \quad\text{as }\rho\to\infty.
\]
Therefore, $\cos(U+e)>0$ places the noiseless observation on the correct side of the decision
hyperplane, whereas $\cos(U+e)<0$ places it on the wrong side. Since the conditional error probability is bounded by one, Lebesgue's dominated convergence theorem gives
\begin{equation}
\begin{aligned}
    \lim_{\rho\to\infty}
    P_{\rm e}(\rho)
    &=
    \E\!\left[
        \mathds{1}\{\cos(U+e)<0\}
    \right] \\
    &=
    \Pr\{\cos(U+e)<0\}
    =
    P_{\infty},
\end{aligned}
\label{eq:sep_high_snr_floor_limit}
\end{equation}
To verify the last equality in (13), we can observe that the error event corresponds to the semicircular arc
$
U+e\in\left(\frac{\pi}{2},\frac{3\pi}{2}\right)$. Consequently, using the von Mises density and the change of variable $x=u+e-\pi$, we obtain
\begin{equation}
\begin{aligned}
\Pr\{\cos(U+e)<0\}
&=\frac{1}{2\pi I_0(\kappa)}
\int_{\pi/2-e}^{3\pi/2-e}e^{\kappa\cos u}du\\
&=\frac{1}{2\pi I_0(\kappa)}
\int_{-\pi/2}^{\pi/2}e^{-\kappa\cos(x-e)}dx
\end{aligned}
\end{equation}
Thus, $P_{\infty}$ is the irreducible high-SNR error caused by
phase realizations that reverse the sign of the noiseless detector
projection. Increasing $\rho$ removes the additive contribution of the heterodyne receiver noise but it cannot correct these phase-induced sign reversals. In addition it is independent from the thermal noise.
\end{corollary}
\begin{corollary}
%     $P_{\infty}$ cannot be expressed in elementary
% form. It can be though evaluated through the absolutely convergent Fourier-Bessel series \cite{gradshteyn2014table}, as given below
% \begin{equation}
% \begin{aligned}
%     P_{\infty}
%     &=
%     \frac{1}{2}
%     -
%     \frac{\gamma}
%     {\sqrt{2\pi}I_0(\kappa)}
%     \sum_{n=0}^{\infty}
%     \frac{
%         \left(\frac{1}{2}\right)_n
%     }{
%         \left(\frac{3}{2}\right)_n n!
%     }
%     \left(
%         -\frac{\gamma^2}{2}
%     \right)^n \\
%     &\hspace{2.8cm}\times
%     \frac{d^{2n+1}}
%     {d\kappa^{2n+1}}
%     I_0(\kappa),
%     \label{eq:heterodyne_oracle_sep_series}
% \end{aligned}
% \end{equation}
% We note that a convergence analysis of the infinite series and an analytic proof are not provided, since the proof can follow directly from \cite[Sec.~8.341]{gradshteyn2014table}. Nevertheless, wherever $P_{\infty}$ is plotted in the numerical results section,  \eqref{eq:heterodyne_oracle_sep_series} is used.
% \end{corollary}
The high-SNR error floor \(P_{\infty}\) does not generally
admit an elementary closed form, but, using the Fourier-Bessel
expansion of the exponential function
\cite{gradshteyn2014table}, it can be expressed as
\begin{equation}
    P_{\infty}
    =
    \frac{1}{2}
    -
    \frac{2}{\pi I_0(\kappa)}
    \sum_{n=0}^{\infty}
    \frac{(-1)^n I_{2n+1}(\kappa)}{2n+1}
    \cos\!\big((2n+1)e\big),
    \label{eq:heterodyne_oracle_sep_series}
\end{equation}
where \(I_m(\cdot)\) denotes the modified Bessel function
of the first kind and order \(m\). To verify \eqref{eq:heterodyne_oracle_sep_series}, consider
the Fourier--Bessel expansion
\[
    \e^{-\kappa\cos(x-e)}
    =
    I_0(\kappa)
    +
    2\sum_{m=1}^{\infty}
    (-1)^m I_m(\kappa)
    \cos\!\big(m(x-e)\big).
\]
For every finite \(\kappa\geq0\), this expansion is absolutely
and uniformly convergent since
\[
    \sum_{m=1}^{\infty}I_m(\kappa)
    =
    \frac{\e^\kappa-I_0(\kappa)}{2}<\infty.
\]
It can therefore be integrated term by term, and since, 
\[
    \int_{-\pi/2}^{\pi/2}
    \cos\!\big(m(x-e)\big)\,dx
    =
    \frac{2}{m}
    \sin\!\left(\frac{m\pi}{2}\right)\cos(me),
\]
all terms of even order vanish, whereas
\(\sin((2n+1)\pi/2)=(-1)^n\), which yields
\eqref{eq:heterodyne_oracle_sep_series}. We note that the same bound
also establishes the absolute convergence of $P_\infty$.
\end{corollary}
\section{CRB Analysis of the Heterodyne Detector}

Next, we characterize the ability of the heterodyne receiver to acquire the
unknown mean phase \(\Delta\), while the concentration parameter
\(\kappa>0\) is treated as known\footnote{ An extension to the case of jointly estimating $\Delta$ and $\kappa$ can directly follow from our analysis, and is thus not considered in this paper.}. Throughout this section, the transmitted BPSK symbol
\(s\in\{-1,+1\}\) is assumed known to the acquisition processor. Let us define
\begin{equation}
    u_\Delta(\bm y,s;\Delta,\kappa)
    =
    \frac{\partial}{\partial\Delta}
    \log p_{\rm het}(\bm y\mid s;\Delta,\kappa).
    \label{eq:het_score_delta_general}
\end{equation}
Then, the corresponding single-observation Fisher information is given as \cite{FItutorial}
\begin{equation}
\begin{aligned}
    J_{\Delta}^{\rm het}
    &=
    \frac{1}{2}
    \sum_{s=\pm1}
    \int_{\mathbb R^2}
    p_{\rm het}(\bm y\mid s;\Delta,\kappa)
        u_\Delta^2(\bm y,s;\Delta,\kappa)
    d\bm y.
\end{aligned}
\label{eq:het_fim_delta_cartesian}
\end{equation}
To analyze the FI, we first need to express the likelihood of the heterodyne receiver output in closed form. This is given in the next proposition.

\begin{proposition}[Phase-marginalized heterodyne likelihood]
\label{prop:het_marginal_likelihood}
Let $    \bm y
    =
    \begin{bmatrix}
        y_I,
        y_Q
    \end{bmatrix}^\top
$ and \(r^2=y_I^2+y_Q^2\). The phase-marginalized heterodyne likelihood
is
\begin{equation}
    p_{\rm het}(\bm y\mid s;\Delta,\kappa)
    =
    \frac{
        \exp[-(r^2+\eta N_s)/(2\sigma_{\rm het}^2)]
    }{
        2\pi\sigma_{\rm het}^2I_0(\kappa)
    }
    I_0(R_s),
    \label{eq:het_likelihood_main}
\end{equation}
where
\begin{equation}
\begin{aligned}
    R_s^2
    &=
    \kappa^2
    +
    \frac{\eta N_s}{\sigma_{\rm het}^4}r^2 +
    \frac{2s\kappa\sqrt{\eta N_s}}{\sigma_{\rm het}^2}
    \left(
        y_I\cos\Delta+y_Q\sin\Delta
    \right).
\end{aligned}
\label{eq:Rs_cartesian_form}
\end{equation}
\end{proposition}

\begin{IEEEproof}
Conditioned on \(\Phi=\phi\), and considering that the heterodyne output follows a bivariate Gaussian distribution we have
\begin{equation}
    \begin{aligned}
          &p(\bm y\mid s,\phi)
    = \\
    &\frac{1}{2\pi\sigma_{\rm het}^2}
    \exp\!\left[
        -\frac{
            (y_I\!-s\sqrt{\eta N_s}\cos\phi)^2\!+\!
            (y_Q\!-\!s\sqrt{\eta N_s}\sin\phi)^2
        }{
            2\sigma_{\rm het}^2
        }
    \right] .
    \end{aligned}
\end{equation}
After expanding the exponent, averaging over
\eqref{eq:von_mises_phase} and by using the Harmonic Addition Theorem \cite{arekat2020fundamental} for the sinusoidal functions, the remaining angular integral, according to \cite{gradshteyn2014table}, admits the
form
\[
    \int_{-\pi}^{\pi}
    \exp(v_{s,I}\cos\phi+v_{s,Q}\sin\phi)d\phi
    =
    2\pi I_0\!\left(
        \sqrt{v_{s,I}^2+v_{s,Q}^2}
    \right),
\]
where
\[
    v_{s,I}
    =
    \kappa\cos\Delta+
    \frac{s\sqrt{\eta N_s}}{\sigma_{\rm het}^2}y_I,
    \qquad
    v_{s,Q}
    =
    \kappa\sin\Delta+
    \frac{s\sqrt{\eta N_s}}{\sigma_{\rm het}^2}y_Q.
\]
Since \(v_{s,I}^2+v_{s,Q}^2=R_s^2\), the result follows.
\end{IEEEproof}

Next, based on Proposition \ref{prop:het_marginal_likelihood} it is straightforward to calculate the score function $u_\Delta$. Let \(q(x)=I_1(x)/I_0(x)\). By differentiating
\eqref{eq:Rs_cartesian_form} we get
\[
    \frac{\partial R_s}{\partial\Delta}
    =
    \frac{s\kappa\sqrt{\eta N_s}}
    {\sigma_{\rm het}^2R_s}
    \left(
        -y_I\sin\Delta+y_Q\cos\Delta
    \right).
\]
Hence, the mean-phase score is
\begin{equation}
\begin{aligned}
    u_\Delta(\bm y,s;\Delta,\kappa)
    &=
    q(R_s)
    \frac{s\kappa\sqrt{\eta N_s}}
    {\sigma_{\rm het}^2R_s}
    \left(
        -y_I\sin\Delta+y_Q\cos\Delta
    \right).
\end{aligned}
\label{eq:score_delta_main}
\end{equation}

\begin{proposition}[Heterodyne mean-phase Fisher information]
\label{prop:heterodyne_fim}
Let \(\rho=\eta N_s/\sigma_{\rm het}^2\). The
FI of \(\Delta\) for a single observation is
\begin{equation}
    J_{\Delta}^{\rm het}
    =
    \frac{\kappa^2\e^{-\rho/2}}
    {2\pi I_0(\kappa)\rho}
    \int_0^\infty
    \lambda^3\e^{-\lambda^2/(2\rho)}
    \mathcal A_\Delta(\lambda;\kappa)
    \,d\lambda,
    \label{eq:jdd_dimless_main}
\end{equation}
where
\begin{equation}
    \mathcal A_\Delta(\lambda;\kappa)
    =
    \int_{-\pi}^{\pi}
    \frac{I_1^2(R)}{I_0(R)}
    \frac{\sin^2\delta}{R^2}
    \,d\delta,
    \label{eq:A_delta_main}
\end{equation}
and
\begin{equation}
    R=
    \sqrt{
        \kappa^2+\lambda^2
        +2\kappa\lambda\cos\delta
    }.
    \label{eq:R_delta_main}
\end{equation}
\end{proposition}

\begin{IEEEproof}
See Appendix~\ref{app:proof_heterodyne_fim}.
\end{IEEEproof}

Based on these propositions, the corresponding CRB  is given as $ \CRB_\Delta^{\rm het} = \frac{1}{J_{\Delta}^{\rm het}}$. However, the exact FI in \eqref{eq:jdd_dimless_main} can be calculated only using numerical integration, thus it does not facilitate further analysis for the investigated system model, neither gives insights on the CRB performance.

\subsection{Information-Theoretic Mean-Phase CRB Bounds}
As such, next, we derive two closed-form information
benchmarks that bound the heterodyne FI  from above and below.
Let us define the following function 
\begin{equation}
    A_m(\kappa)
    =
    \frac{I_m(\kappa)}{I_0(\kappa)},
    \qquad m\in\mathbb N,
    \label{eq:Am_definition}
\end{equation}
for the von Mises trigonometric moments. Then the following Propositions hold.
\begin{proposition}[Complete-phase information bound]
\label{prop:cpb}
If the latent phase $\Phi$ were to be observed directly, the FI of $\Delta$ would be given as
\begin{equation}
    J_{\Phi,\Delta}
    =
    \frac{\kappa^2}{2}[1-A_2(\kappa)].
    \label{eq:Jphi_main}
\end{equation}
Moreover,
\begin{equation}
    J_{\Delta}^{\rm het}
    \leq
    J_{\Phi,\Delta}.
    \label{eq:complete_phase_delta_order}
\end{equation}
\end{proposition}

\begin{IEEEproof}
See Appendix~\ref{app:cpb}.
\end{IEEEproof}

This proposition, gives the best mean-phase acquisition performance if
the random phase itself were revealed. It is independent of $N_s$
because optical energy no longer matters once the latent phase shift 
is directly observed. The bound becomes tight only when the heterodyne observation
is sufficiently informative about that phase sample, i.e., when the number of photons increases towards infinity.

\begin{proposition}[Moment-information lower bound]
\label{prop:moment}
Let us again define $\rho=\eta N_s/\sigma_{\mathrm{het}}^2$, then the heterodyne mean-phase Fisher information satisfies
\begin{equation}
    J_{\Delta}^{\rm het}
    \geq
    J_\Delta^{\rm mom},
    \label{eq:moment_delta_order}
\end{equation}
where
\begin{equation}
    J_\Delta^{\rm mom}
    =
    \frac{\rho A_1^2(\kappa)}
    {1+\frac{\rho}{2}[1-A_2(\kappa)]}.
    \label{eq:Jmom_main}
\end{equation}
\end{proposition}

\begin{IEEEproof}
See Appendix~\ref{app:moment}.
\end{IEEEproof}

The moment bound captures the information visible from the first two
observable moments only. It has the form $cN_s/(1+dN_s)$, thus it grows
linearly at low photon number and saturates as the photon number increases. Finally, combining the two propositions we can bound the CRB of the mean phase shift of the dephasing channel as
\begin{equation*}
    \frac{2}{\kappa^2[1-A_2(\kappa)]}
    \leq
    \CRB_\Delta^{\rm het}
    \leq
    \frac{1+\frac{\rho}{2}[1-A_2(\kappa)]}
    {\rho A_1^2(\kappa)}.
\end{equation*}
Using \(1-A_2(\kappa)=2A_1(\kappa)/\kappa\), the CRB bound 
 can equivalently be written as
\begin{equation}
    \frac{1}{\kappa A_1(\kappa)}
    \leq
    \CRB_\Delta^{\rm het}
    \leq
    \frac{1}{\kappa A_1(\kappa)}
    +
    \frac{1}{\rho A_1^2(\kappa)}.
    \label{eq:delta_crb_bounds}
\end{equation}
This form separates the ultimate dephasing sensing floor from the finite-SNR
penalty of the heterodyne receiver. The latter decreases as \(1/\rho\),
while the former depends only on the concentration of the random
phase and cannot be reduced by increasing the photon number. Hence,
the two bounds coincide as \(\rho\to\infty\), showing that the heterodyne
detection asymptotically recovers all the mean-phase information
available, since its CRB becomes equal to the complete phase CRB. Thus, heterodyne receiver is asymptotically optimal, which shall be also verified when we calculate the quantum CRB.  

\section{Quantum Fisher Information Benchmark}
\label{sec:qfi_benchmark}
To characterize the phase information contained in the received
optical state before a measurement is selected, we also consider the
QFI for the mean phase \(\Delta\). The
following analysis provides insight into how closely the heterodyne
receiver approaches the optimal sensing performance permitted by the
given dephasing and thermal channel. Moreover, the tools and
calculation method developed below can also be applied to other
quantum channels and different states in the transmission phase, like squeezed states. As in the preceding sensing analysis, the
transmitted BPSK symbol \(s\) is known, the concentration
\(\kappa>0\) is fixed, and only \(\Delta\) is estimated. We first derive the Fock basis representation of the received state.
Since \(\hat n|n\rangle=n|n\rangle\), where $\hat n$ is the annihilation operator, the Fock basis elements of the
conditional state in \eqref{eq:qfi_received_state} are
\begin{align}
    [\rho_{\mathrm{out}}]_{mn}
    &=
    \langle m|
    U_\phi\sigma_sU_\phi^\dagger
    |n\rangle =
    [\sigma_s]_{mn}
    \e^{j(m-n)\phi},
    \label{eq:qfi_conditional_fock}
\end{align}
with $
    \sigma_s
    =
    D\!\left(s\sqrt{\eta N_s}\right)
    \rho_{\rm th}(N_a)
    D^\dagger\!\left(s\sqrt{\eta N_s}\right)$, $U_\phi=\e^{j\phi \hat{n}}$ and
\begin{equation*}
    \rho_{\rm th}(N_a)
    =
    \sum_{r=0}^{\infty}
    \frac{N_a^r}{(N_a+1)^{r+1}}
    |r\rangle\langle r|,
\end{equation*}
where $\rho_{\rm th}$ is the density operator of the thermal state \cite{aspachs2009phase}. Due to symmetry, we only need to calculate the case of $s=+1$, so in the rest of the analysis, $s=+1$ is assumed.     
    Substitution into \eqref{eq:qfi_received_state} yields
\begin{align}
    [\rho_\mathrm{out}]_{mn}
    &=
    [\sigma_{+1}]_{mn}
    \int_{-\pi}^{\pi}
    p_\Phi(\phi;\Delta,\kappa)
    \e^{j(m-n)\phi}\,d\phi.
    \label{eq:qfi_vm_integral}
\end{align}
The circular moments of the von Mises distribution satisfy
\begin{equation*}
    \int_{-\pi}^{\pi}
    p_\Phi(\phi;\Delta,\kappa)
    \e^{j\ell\phi}\,d\phi
    =
    A_\ell(\kappa)\e^{j\ell\Delta}.
    \label{eq:qfi_vm_moment}
\end{equation*}
Setting \(\ell=m-n\) gives
\begin{equation}
    [\rho_\mathrm{out}]_{mn}
    =
    [\sigma_{+1}]_{mn}
    A_{m-n}(\kappa)
    \e^{j(m-n)\Delta},
    \label{eq:qfi_thermal_fock_state}
\end{equation}
where \(A_{-\ell}(\kappa)=A_\ell(\kappa)\). Thus, von Mises dephasing suppresses the Fock-basis components of order
\(|m-n|\) by the factor \(A_{m-n}(\kappa)\), while leaving the diagonal populations unchanged since \(A_0(\kappa)=1\), therefore, causing decoherence.

Next, differentiating the phase
factor with respect to \(\Delta\)  gives
\begin{equation}
    [\partial_\Delta\rho_\mathrm{out}]_{mn}
    =
    j(m-n)[\rho_\mathrm{out}]_{mn}.
    \label{eq:qfi_thermal_derivative}
\end{equation}
Let us define the eigendecomposition of the phase-averaged state to be
\begin{equation*}
    \rho_\mathrm{out}
    =
    \sum_a p_a|a\rangle\langle a|.
    \label{eq:qfi_state_eigendecomposition}
\end{equation*}
Here, the eigenvalues $p_a$ are a function of $\eta N_s,N_a,\kappa$ and they are independent of \(\Delta\), which follows immediately from \eqref{eq:qfi_conditional_fock}. Then the exact QFI for estimating \(\Delta\) is then given by
\cite{liu2020quantum} as
\begin{equation}
    J_{\mathrm Q,\Delta}
    =
    2
    \!\!\!\sum_{\substack{a,b\\p_a+p_b>0}}
    \frac{
        \left|
            \langle a|
            \partial_\Delta\rho_\mathrm{out}
            |b\rangle
        \right|^2
    }{
        p_a+p_b
    }.
    \label{eq:qfi_thermal_spectral}
\end{equation}
% We note that  $[\rho_\mathrm{out}]_{mn}$ and $[\partial_\Delta\rho_\mathrm{out}]_{mn}$ are implicitly used in the calculation of \eqref{eq:qfi_thermal_spectral}, since the value of $p_a$ depends on $[\rho_\mathrm{out}]_{mn}$. 
Equation \eqref{eq:qfi_thermal_spectral} provides the exact
infinite dimensional QFI formula. For the numerical evaluation, though, the operators are
restricted to a sufficiently large Fock subspace. Within that
subspace, \(\rho_\mathrm{out}\) and
\(\partial_\Delta\rho_\mathrm{out}\) are constructed from
\eqref{eq:qfi_thermal_fock_state} and
\eqref{eq:qfi_thermal_derivative}, respectively. The density operator
is then diagonalized, and the resulting eigenvalues and eigenvectors
are substituted into \eqref{eq:qfi_thermal_spectral}. The Fock cutoff
is increased until calculation of \(J_{\mathrm Q,\Delta}\) is essentially unchanged.

\subsection{QCRB bounds}
In this subsection, we complement the previous analysis by providing two rigorous bounds on the derived QFI. The exact QFI can be evaluated numerically from its spectral representation, but it does not yield a closed-form dependence on the system parameters. We therefore derive rigorous lower and upper bounds
on the QFI, which produce corresponding upper and lower endpoints for the QCRB. These bounds provide analytically tractable quantum
benchmarks and quantify how tightly the exact sensing limit is localized without requiring numerical diagonalization. Before we proceed, we will introduce two important quantities for our derivations. First, from Proposition 4, the
Fisher information carried by direct observation of the latent phase is given as
    $J_{\Phi,\Delta}
    =
    \frac{\kappa^2}{2}
    [1-A_2(\kappa)]$.
Using the modified-Bessel recurrence
    $1-A_2(\kappa)
    =
    \frac{2A_1(\kappa)}{\kappa}$,
this becomes
\begin{equation}
    J_{\Phi,\Delta}
    =
    \kappa A_1(\kappa).
    \label{eq:qfi_latent_phase}
\end{equation}
Second, the  QFI for phase estimation of a displaced thermal state under a deterministic
phase rotation is given as
\begin{equation}
    J_{\mathrm{dth}}
    =
    \frac{4\eta N_s}{2N_a+1},
    \label{eq:qfi_displaced_thermal}
\end{equation}
as established for phase estimation with displaced thermal
states in \cite{aspachs2009phase,Pinel2013}. Then, the bounds for the QCRB can be derived based on the following Propositions.

\begin{proposition}[Extended-convexity bound]
\label{prop:qfi_upper_bound}
The QFI satisfies
\begin{equation}
    J_{\mathrm Q,\Delta}
    \le
    J_{\mathrm Q,\Delta}^{\mathrm U}
    =
    \left[
        \frac{1}{\kappa A_1(\kappa)}
        +
        \frac{2N_a+1}{4\eta N_s}
    \right]^{-1}.
    \label{eq:qfi_upper_bound}
\end{equation}
Consequently,
\begin{equation*}
    C_{\mathrm Q,\Delta}
    \ge
    C_{\mathrm{EC},\Delta},
    \label{eq:qcrb_ec_inequality}
\end{equation*}
where
\begin{equation}
    C_{\mathrm{EC},\Delta}
    =
        \frac{1}{\kappa A_1(\kappa)}
        +
        \frac{2N_a+1}{4\eta N_s}.
    \label{eq:qcrb_extended_convexity}
\end{equation}
\end{proposition}

\begin{IEEEproof}
The Proof can be found in Appendix \ref{prop:qcrb_upper}.
\end{IEEEproof}
Next, the Proposition for the QFI lower bound is presented. 
\begin{proposition}[Quantum quadrature-moment bound]
\label{prop:qfi_lower_bound}
The QFI satisfies
\begin{equation}
    J_{\mathrm Q,\Delta}
    \ge
    J_{\mathrm Q,\Delta}^{\mathrm L}
    =
    \left[
        \frac{1}{\kappa A_1(\kappa)}
        +
        \frac{2N_a+1}{
            4\eta N_s A_1^2(\kappa)
        }
    \right]^{-1}.
    \label{eq:qfi_lower_bound}
\end{equation}
Consequently,
\begin{equation*}
    C_{\mathrm Q,\Delta}
    \le
    C_{\mathrm{QM},\Delta},
    \label{eq:qcrb_qm_inequality}
\end{equation*}
where
\begin{equation}
    C_{\mathrm{QM},\Delta}
    =
        \frac{1}{\kappa A_1(\kappa)}
        +
        \frac{2N_a+1}{
            4\eta N_s A_1^2(\kappa)
        }.
    \label{eq:qcrb_quadrature_moment}
\end{equation}
\end{proposition}

\begin{IEEEproof}
Please refer to Appendix \ref{app:qfi_lower_bound}.
\end{IEEEproof}
Based on these bounds QCRB is bounded as
\begin{equation*}
 \frac{1}{\kappa A_1(\kappa)}
        +
        \frac{2N_a+1}{4\eta N_s} \le  C_{\mathrm Q,\Delta}
    \le
    \frac{1}{\kappa A_1(\kappa)} +\frac{2N_a+1}{
            4\eta N_s A_1^2(\kappa)}.
\end{equation*}
Several interesting points that arise from these bounds are the following. First, both bounds converge asymptotically to the common floor
\(C_{\mathrm Q,\Delta}\rightarrow[ M\kappa A_1(\kappa)]^{-1}\), which is caused by the von Mises phase diffusion.  We note that is the same asymptotic bound which appears for the heterodyne CRB in \eqref{eq:delta_crb_bounds}, which verifies that heterodyne receiver is optimal as the number of photons increases.  Second, the thermal photon number \(N_a\) enters through the
finite-energy term \((2N_a+1)/(4\eta N_s)\). Thermal noise
therefore raises the QCRB at finite energy and slows convergence to the
phase-diffusion floor, but it does not change the QCRB floor itself for
fixed \(\kappa\).  Finally, the interval width between the upper and the lower bound is
\(C_{\mathrm{QM},\Delta}-C_{\mathrm{EC},\Delta}
=(2N_a+1)[A_1^{-2}(\kappa)-1]/(4\eta N_s)\). Hence, the bounds
become tight as \(\mathcal{O}(\eta N_s^{-1})\) and are especially
close for larger \(\kappa\). In the deterministic-phase limit
\(\kappa\rightarrow\infty\), they coincide with the exact result
\(C_{\mathrm Q,\Delta}=(2N_a+1)/(4M\eta N_s)\). Conversely, as
\(\kappa\rightarrow0\), the QCRB diverges because a uniform phase
distribution contains no information about its mean direction
\(\Delta\).

\section{Closed-Form Photon-Allocation Limits}
\label{sec:photon_allocation}
To showcase the significance, and potential applications, of the preceding analysis, let us consider a sequential sensing and communication scenario under a fixed photon budget. The transmitter first sends known BPSK acquisition symbols, which the receiver uses to estimate the mean phase \(\Delta\), and subsequently sends BPSK data symbols that are detected using the acquired phase information. This pilot-assisted structure is consistent with current literature \cite{Zhu2023TimeDivision,Zhang2022TimeDivisionISAC} and with our preceding sensing analysis, since the  CRBs for \(\Delta\) were derived conditionally on the BPSK symbol \(s\). 

Let \(N_{\rm tot}\) denote the total photon budget and \(n\) the total number of channel uses available. We introduce \(\alpha\in(0,1)\) as the fraction of the photon budget allocated to phase acquisition and \(\beta\in(0,1)\) as the fraction of channel uses assigned to this task. Accordingly,
\begin{equation*}
\begin{aligned}
    N_\Delta &= \alpha N_{\rm tot},
    &\qquad
    N_{\rm c} &= (1-\alpha)N_{\rm tot},\\
    n_\Delta &= \beta n,
    &\qquad
    n_{\rm c} &= (1-\beta)n,
\end{aligned}
\label{eq:tdma_resource_allocation}
\end{equation*}
where \(\beta n\) and \((1-\beta)n\) are assumed to be integers. Then, defining    $\rho_{\rm tot} \triangleq \frac{\eta N_{\rm tot}}{\sigma_{\rm het}^{2}}$, the per-symbol heterodyne SNRs in the two stages become
\begin{equation*}
    \rho_{\Delta}
    =
    \frac{\alpha\rho_{\rm tot}}{\beta n},
    \qquad
    \rho_{{\rm c}}
    =
    \frac{(1-\alpha)\rho_{\rm tot}}{(1-\beta)n}.
    \label{eq:tdma_stage_snrs}
\end{equation*}
This model allows the photon and channel uses to be assigned to the two tasks independently. For a fixed pair \((\alpha,\beta)\), increasing \(n\) provides more known observations for estimating \(\Delta\), since \(n_\Delta=\beta n\), but distributes the available photons over a larger number of symbols and therefore reduces the individual SNR per symbol. Consequently, the sensing requirement generally imposes a minimum total number of channel uses, whereas the communication SEP requirement imposes a maximum number of channel uses.

Let \(\tau_\Delta\) to denote the  phase estimation target and let \(\epsilon_{\rm SEP}\) denote the communication SEP requirement. For a given resource allocation \((\alpha,\beta)\), the exact feasible set of total channel uses is described as
\begin{equation}
\begin{aligned}
    \mathcal N^{\rm feas}(\alpha,\beta)
    =
    \Bigg\{
        n\in\mathbb N:
        \;&
        \CRB_\Delta^{(\beta n)}
        \left(
            \frac{\alpha\rho_{\rm tot}}{\beta n},
            \kappa
        \right)
        \leq \tau_\Delta,
        \\
        &
        P_{\rm e}
        \left(
            \frac{(1-\alpha)\rho_{\rm tot}}
                 {(1-\beta)n}
        \right)
        \leq \epsilon_{\rm SEP}
    \Bigg\}.
\end{aligned}
\label{eq:exact_tdma_feasible_set}
\end{equation}
The first condition accounts for the \(\beta n\) symbols that are used for the phase acquisition, while the second characterizes the reliability of the \((1-\beta)n\) communication symbols. Direct characterization of \(\mathcal N^{\rm feas}(\alpha,\beta)\) requires repeated numerical evaluation and inversion of the exact heterodyne Fisher information and SEP. The preceding analytical results can be instead  used to determine this region explicitly. In particular, the mean-phase CRB bound provides a sufficient lower limit \(n_\Delta^{\rm suf}(\alpha,\beta)\) on the total number of channel uses, while the SEP approximation provides an upper limit \(n_{\rm SEP}^{\rm LG,int}(\alpha,\beta)\). The feasible set of total channel uses can be written as 
\begin{equation}
    \mathcal N_{\rm pred}^{\rm feas}(\alpha,\beta)
    =
    \left\{
        n\in\mathbb N:
        n_\Delta^{\rm suf}(\alpha,\beta)
        \leq n
        \leq
        n_{\rm SEP}^{\rm LG,int}(\alpha,\beta)
    \right\},
    \label{eq:predicted_tdma_feasible_set}
\end{equation}
which is nonempty whenever
\begin{equation}
    n_\Delta^{\rm suf}(\alpha,\beta)
    \leq
    n_{\rm SEP}^{\rm LG,int}(\alpha,\beta).
    \label{eq:predicted_tdma_joint_feasibility}
\end{equation}
Values below the lower endpoint do not provide sufficiently many acquisition observations to guarantee the phase-estimation accuracy. In contrast, values above the upper endpoint distribute the communication photon budget over too many data symbols and violate the SEP constraint. The interval between the two limits therefore identifies the total numbers of channel uses for which both requirements are predicted to be simultaneously feasible under the selected photon and channel-use allocation. Selecting a particular operating point within this interval, or optimizing \(\alpha\) and \(\beta\), would require an additional objective or constraint, such as throughput, and is beyond the scope of the present analysis. The following subsections use the previously derived CRB bound and SEP approximation to obtain the two endpoints of \eqref{eq:predicted_tdma_joint_feasibility} in closed form, for a given set of $(\alpha,\beta)$, which then characterizes the chosen SEP and CRB constraints as jointly feasible or not.

\subsection{Mean-Phase Acquisition Lower Limits}
Here, we give the lower number of photons that are needed for the CRB constraint to be guaranteed. The sensing stage contains \(n_\Delta=\beta n\) independent heterodyne observations, each with SNR of $\rho_{\Delta}$. Since Fisher information is additive, the corresponding CRB is
\begin{equation*}
    \CRB_{\Delta}^{(\beta n)}
    =
    \frac{1}{
        \beta n
        J_{\Delta}^{\rm het}
        \left(
            \rho_{\Delta},\kappa
        \right)
    }.
    \label{eq:fixed_budget_delta_crb}
\end{equation*}
Substituting
\(\rho_{\Delta}=\alpha\rho_{\rm tot}/(\beta n)\)
into the moment-information upper CRB bound in
\eqref{eq:delta_crb_bounds} gives
\begin{equation}
\begin{aligned}
    \CRB_{\Delta}^{(\beta n)}
    &\leq
    \overline{\CRB}_{\Delta}(n;\alpha,\beta)
    \\
    &=
    \frac{1}{
        \alpha\rho_{\rm tot}A_1^2(\kappa)
    }
    +
    \frac{
        1-A_2(\kappa)
    }{
        2\beta n A_1^2(\kappa)
    }.
\end{aligned}
\label{eq:crb_delta_fixed_budget_bound}
\end{equation}
Requiring \(\overline{\CRB}_{\Delta}(n;\alpha,\beta)\leq\tau_\Delta\)
gives the sufficient lower limit on the total number of channel uses
\begin{equation}
    n_\Delta^{\rm suf}(\alpha,\beta)
    =
    \left\lceil
        \frac{
            1-A_2(\kappa)
        }{
            2\beta
            \left[
                A_1^2(\kappa)\tau_\Delta
                -
                (\alpha\rho_{\rm tot})^{-1}
            \right]
        }
    \right\rceil,
    \label{eq:n_delta_sufficient}
\end{equation}
provided that
\begin{equation}
    \tau_\Delta
    >
    \frac{1}{
        \alpha\rho_{\rm tot}A_1^2(\kappa)
    }.
    \label{eq:n_delta_sufficient_feasibility}
\end{equation}
The condition in \eqref{eq:n_delta_sufficient_feasibility} follows from the \(n\)-independent term in \eqref{eq:crb_delta_fixed_budget_bound}. Increasing the total number of channel uses reduces only the second term, which accounts for averaging the random phase fluctuations over the \(\beta n\) acquisition observations. The first term,
\(1/[\alpha\rho_{\rm tot}A_1^2(\kappa)]\), cannot be reduced by increasing \(n\); for fixed \(\kappa\), it can be lowered only by allocating a larger fraction \(\alpha\) of the photon budget to sensing or by increasing the total available photon budget. On the other hand, the allocation factor \(\beta\) controls how quickly the second term decreases with \(n\), but does not affect this fixed-budget floor.

\subsection{Communication-Limited Upper Allocation}
Next, we provide the analysis for the number of photons allocation so that the SEP threshold holds.  We note that the two-point GL approximation in
\eqref{eq:sep_two_point_gl} contains two \(Q(\cdot)\) terms weighted by
\(\Gamma_1\) and \(\Gamma_2\) and evaluated using two different
coefficients \(c_1\) and \(c_2\). Since \(c_1\neq c_2\), setting this
expression equal to a prescribed SEP target does not admit a direct
inversion through a single inverse error function. Thus, to obtain an
upper allocation-limit formula, we introduce the following effective
approximation
\begin{equation}
    \Gamma_1Q(c_1\sqrt{\rho})
    +
    \Gamma_2Q(c_2\sqrt{\rho})
    \approx
    \Gamma_{\rm eff}Q(c_{\rm eff}\sqrt{\rho}).
    \label{eq:two_point_collapse_definition}
\end{equation}
Using the Maclaurin series of the error function we get
\begin{equation*}
    Q(c\sqrt{\rho})
    =
    \frac{1}{2}
    -
    \frac{c}{\sqrt{2\pi}}\sqrt{\rho}
    +
    O(\rho^{3/2}),
    \label{eq:Q_small_t_expansion}
\end{equation*}
and by ignoring the higher-order terms, we obtain
\begin{equation*}
\begin{aligned}
    &\Gamma_1Q(c_1\sqrt{\rho})
    +
    \Gamma_2Q(c_2\sqrt{\rho})
=
    \frac{\Gamma_1+\Gamma_2}{2}
    -
    \frac{
        \Gamma_1c_1+\Gamma_2c_2
    }{
        \sqrt{2\pi}
    }\sqrt{\rho}.
\end{aligned}
\label{eq:two_point_small_t_expansion}
\end{equation*}
Matching this expansion with the right-hand side of
\eqref{eq:two_point_collapse_definition} gives
\begin{equation*}
    \Gamma_{\rm eff}
    =
    \Gamma_1+\Gamma_2,
    \qquad
    c_{\rm eff}
    =
    \frac{
        \Gamma_1c_1+\Gamma_2c_2
    }{
        \Gamma_1+\Gamma_2
    }.
    \label{eq:effective_sep_parameters}
\end{equation*}
Therefore, the derived auxiliary approximation is
\begin{equation}
    \widetilde P_{\rm e}(\rho)
    =
    P_\infty
    +
    \Gamma_{\rm eff}
    Q(c_{\rm eff}\sqrt{\rho}).
    \label{eq:sep_two_point_collapsed}
\end{equation}
We note that the original two-term expression derived in
\eqref{eq:sep_two_point_gl} remains the approximation used for the
SEP evaluation and numerical curves throughout our paper. Equation
\eqref{eq:sep_two_point_collapsed} is introduced only to obtain an
explicit allocation formula. Let
\begin{equation}
    \delta_{\rm SEP}
    =
    \epsilon_{\rm SEP}-P_\infty.
    \label{eq:delta_sep_definition}
\end{equation}

The communication stage is allocated \((1-\alpha)N_{\rm tot}\)
photons and \((1-\beta)n\) channel uses. Its per-symbol SNR is therefore
\begin{equation}
    \rho_{{\rm c}}
    =
    \frac{
        (1-\alpha)\rho_{\rm tot}
    }{
        (1-\beta)n
    }.
    \label{eq:communication_stage_snr}
\end{equation}
Substituting \(\rho=\rho_{{\rm c}}\) into
\eqref{eq:sep_two_point_collapsed}, setting the result equal to
\(\epsilon_{\rm SEP}\), and using
\(Q(z)=\frac{1}{2}\erfc(z/\sqrt{2})\) gives
\begin{equation*}
    \frac{
        2\delta_{\rm SEP}
    }{
        \Gamma_{\rm eff}
    }
    =
    \erfc\!\left(
        c_{\rm eff}
        \sqrt{
            \frac{
                (1-\alpha)\rho_{\rm tot}
            }{
                2(1-\beta)n
            }
        }
    \right).
    \label{eq:two_point_n_equation_erfc}
\end{equation*}
Finally, applying \(\erfc^{-1}(\cdot)\) and solving for the total
number of channel uses \(n\) yields
\begin{equation}
    n_{\rm SEP}^{\rm LG}(\alpha,\beta)
    =
    \frac{
        (1-\alpha)\rho_{\rm tot}c_{\rm eff}^2
    }{
        2(1-\beta)
        \left[
            \erfc^{-1}\!\left(
                \dfrac{
                    2(\epsilon_{\rm SEP}-P_\infty)
                }{
                    \Gamma_{\rm eff}
                }
            \right)
        \right]^2
    }.
    \label{eq:n_sep_two_point_closed_form}
\end{equation}
For this inversion, the following condition must hold:
\begin{equation*}
    0
    <
    \frac{
        2(\epsilon_{\rm SEP}-P_\infty)
    }{
        \Gamma_{\rm eff}
    }
    <
    1.
    \label{eq:two_point_n_feasibility}
\end{equation*}
The lower inequality requires
\(\epsilon_{\rm SEP}>P_\infty\); otherwise, the requested SEP lies at
or below the irreducible phase-error floor. The upper inequality
ensures that the target lies below the zero-SNR value of the collapsed
approximation and therefore produces a finite upper allocation limit. We note that \(n_{\rm SEP}^{\rm LG}(\alpha,\beta)\) is a closed-form
prediction of the communication-limited upper allocation obtained from
the effective-tail collapse. It is not a rigorous upper or lower bound
on the crossing of the original two-term approximation in
\eqref{eq:sep_two_point_gl}, thus its accuracy will be evaluated numerically.

\section{Numerical Results}
\label{sec:numerical_results}

In this section we evaluate the communication and mean-phase acquisition performance
of the investigated channel. Unless otherwise stated, \(N_a=2\), so
that \(\sigma_{\rm het}^2=(N_a+1)/2=3/2\), and  \(\eta=0.8\). The exact SEP is computed
from \eqref{eq:heterodyne_hyperplane_sep}, its two-point
GL approximation from \eqref{eq:sep_two_point_gl}, and
the exact mean-phase Fisher information by numerical integration of
\eqref{eq:jdd_dimless_main}.

\begin{figure}[!t]
    \centering
    \begin{subfigure}{\columnwidth}
        \centering
        \includegraphics[width=0.92\linewidth]
        {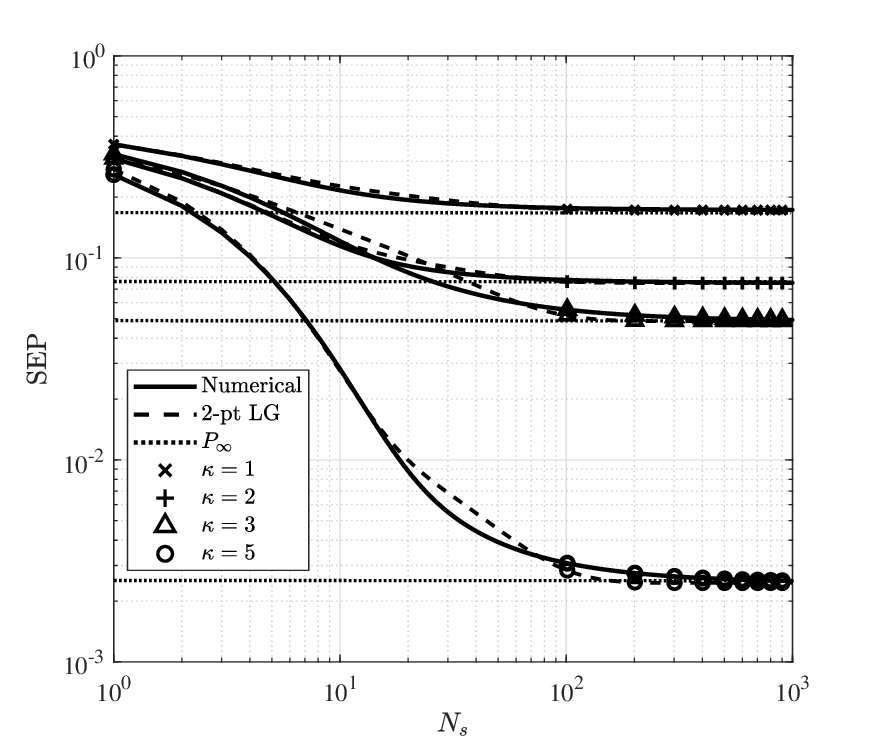}
        \caption{SEP versus the per-symbol photon number \(N_s\).}
        \label{fig:sep_ns}
    \end{subfigure}
    \begin{subfigure}{0.95\columnwidth}
        \centering
        \includegraphics[width=\linewidth]
        {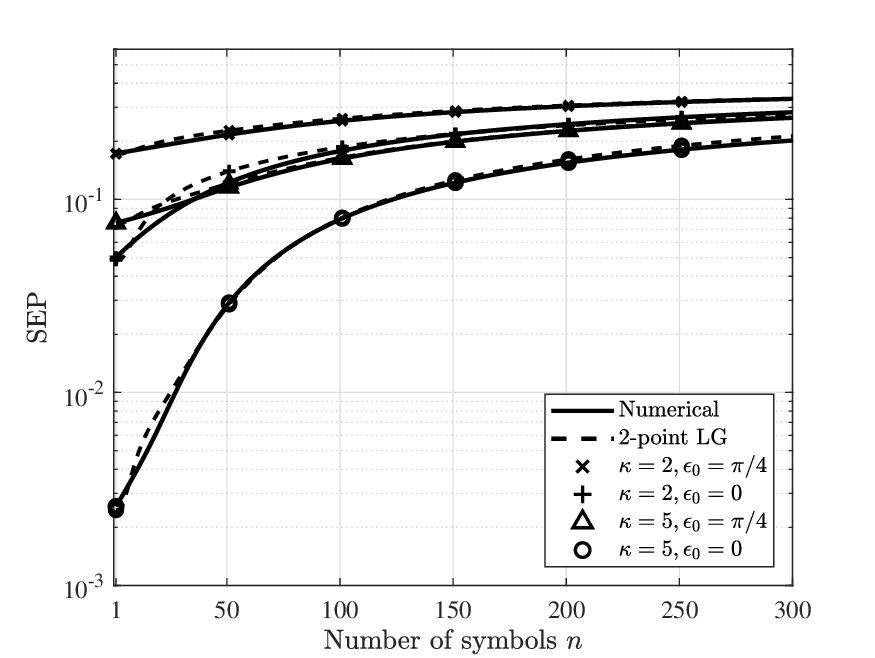}
        \caption{SEP versus \(n\) under \(N_s=N_{\rm tot}/n\).}
        \label{fig:sep_n}
    \end{subfigure}
    \caption{SEP and the two-point
    GL approximation.}
    \label{fig:sep}
\end{figure}

For Fig.~\ref{fig:sep}, we consider
\(e\in\{0,\pi/4\}\) to verify the approximation under both aligned
and mismatched detection. We also consider \(\kappa\in\{2,3,4,5\}\).
Fig.~\ref{fig:sep_ns} compares the exact SEP with the approximation
of Proposition~\ref{prop:two_point_gl_sep}, over a range of transmitted photons. We observe that the two curves remain close
over the complete photon range, showing that the two point
GL approximation effectively represents \eqref{eq:heterodyne_hyperplane_sep}. At small \(N_s\), the SEP approaches \(1/2\)
because heterodyne noise dominates. As \(N_s\) increases, it approaches
the phase-induced floor \(P_\infty\), consistently with
Corollary 1. Furthermore, increasing \(\kappa\) concentrates the
phase around \(\Delta\), thereby reducing the probability of a
phase-induced sign reversal and lowering SEP in the low and high number of photons regime. In Fig.~\ref{fig:sep_n} we consider a total available photon budget of $N_\mathrm{tot}=500$. The available SNR of the transmission is then calculated as $N_s = N_\mathrm{tot}/n$, where $n$ is practically the total number of symbols the transmitter selects to send under the available budget of $N_\mathrm{tot}$. Again, the two point GL approximation can be seen to accurately follow the numerically calculated SEP value. Fig.~\ref{fig:sep_n}, thus, verifies that the designed two point GL approximation is appropriate for the analysis of Section \ref{sec:photon_allocation}.

\begin{figure}[!t]
    \centering
    \begin{subfigure}{\columnwidth}
        \centering
        \includegraphics[width=\linewidth]
        {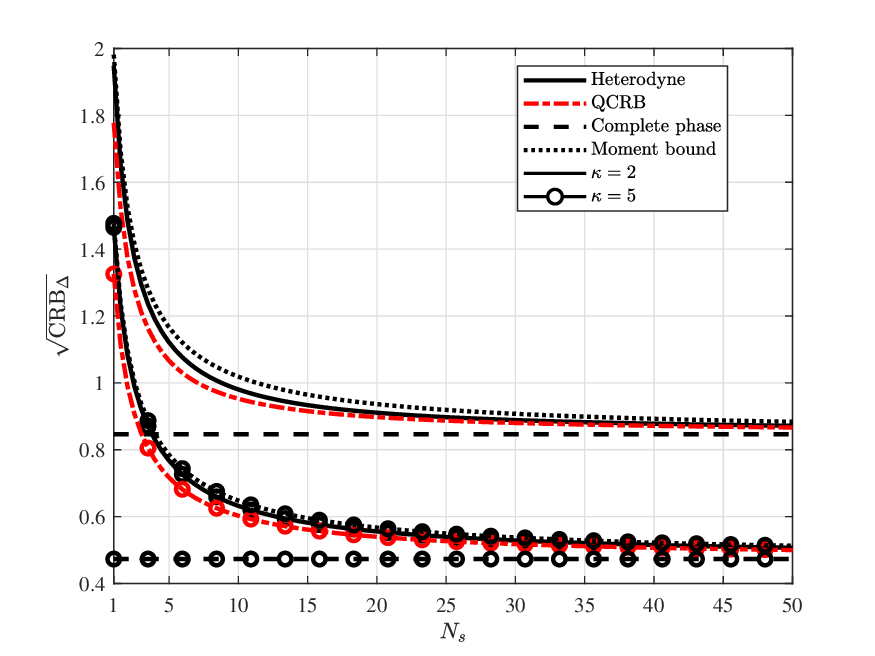}
        \caption{Heterodyne CRB comparison versus \(N_s\).}
        \label{fig:crb_delta_ns}
    \end{subfigure}
    \begin{subfigure}{\columnwidth}
        \centering
        \includegraphics[width=\linewidth]
        {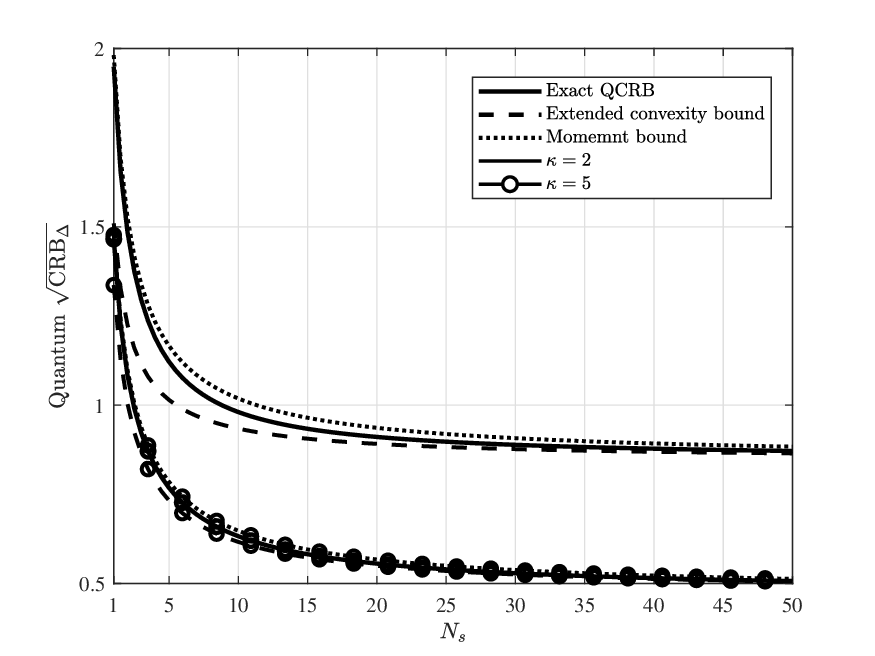}
        \caption{Quantum CRB vs $N_s$.}
        \label{fig:qcrb}
    \end{subfigure}
    \caption{CRB comparisons.}
    \label{fig:crb}
\end{figure}

\begin{figure}[t]
    \centering
    \begin{subfigure}{\columnwidth}
        \centering
        \includegraphics[width=\linewidth]
        {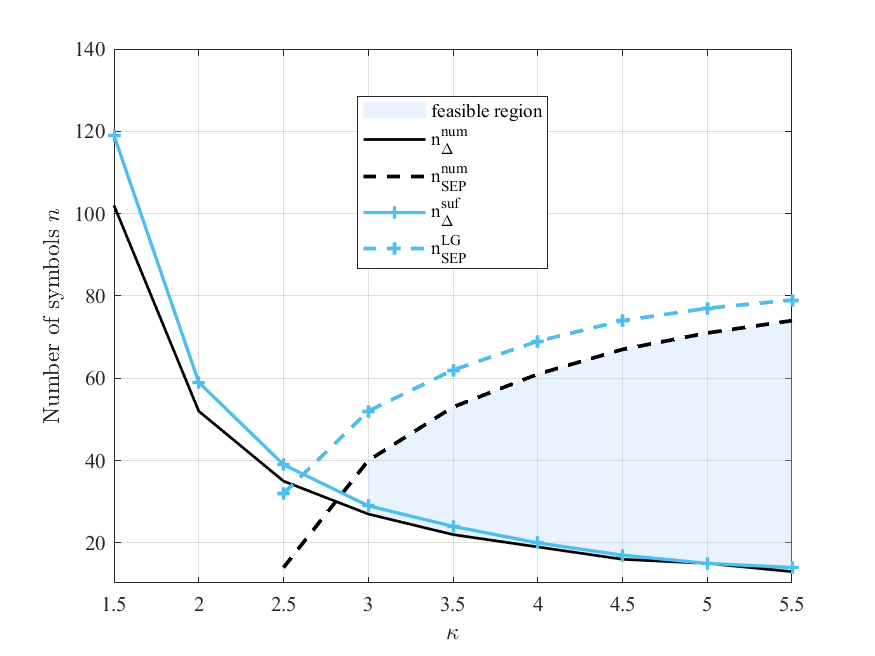}
        \caption{Numerical and closed-form photon-allocation limits vs $\kappa$}
        \label{fig:blocklength_delta1}
    \end{subfigure}
    \begin{subfigure}{\columnwidth}
        \centering
        \includegraphics[width=\linewidth]
        {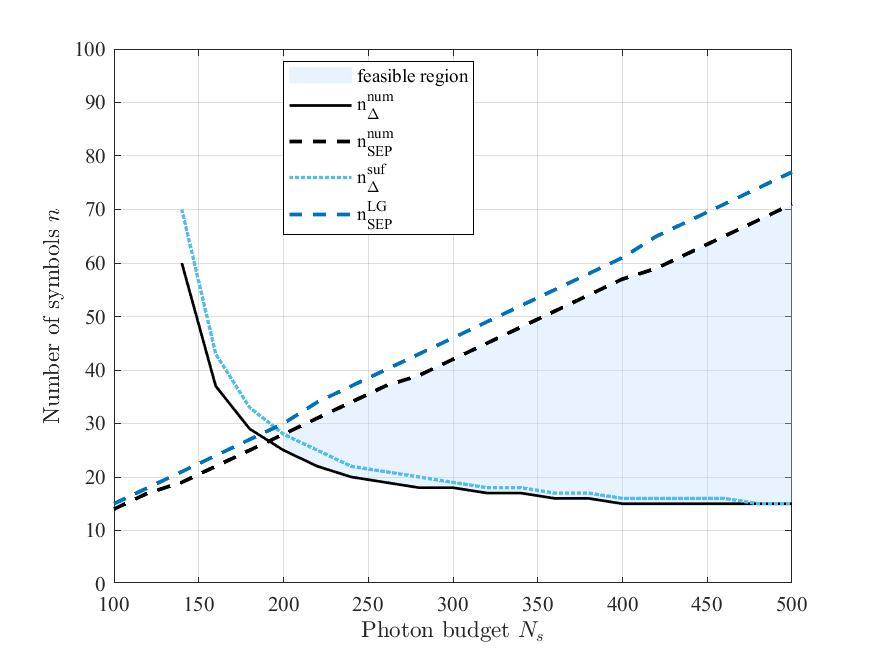}
        \caption{Numerical and closed-form photon-allocation limits vs $N_s$.}
        \label{fig:blocklength_delta2}
    \end{subfigure}
    \caption{Numerical and closed-form photon-allocation limits.}
    \label{fig:blocklength_delta}
\end{figure}

Fig.~\ref{fig:crb_delta_ns} shows the comparison between the exact heterodyne root CRB
obtained from \eqref{eq:jdd_dimless_main} with the complete-phase
benchmark of Proposition~\ref{prop:cpb} and the
moment-information bound of Proposition~\ref{prop:moment}. The exact
root-CRB decreases with \(N_s\) and approaches the complete-phase
benchmark as the number of photons increase and the heterodyne receiver can gather substantial information about the latent phase from it's noisy output. The moment bound remains comparatively close to the exact result, indicating that the first two
Cartesian moments retain substantial information about \(\Delta\). This indicates that a practical algorithm for estimating $\Delta$ needs only the first two moments, therefore, a regression-like estimation should suffice. Moreover, increasing \(\kappa\) improves acquisition because the mean
phase direction becomes more sharply defined. An interesting note is that QCRB is also consistently close to the heterodyne receiver, which indicates that the heterodyne receiver is near-optimal for estimating the mean phase rotation of the studied von Mises thermal channel. In addition, Fig. \ref{fig:qcrb} plots the exact rooted QCRB versus the derived bounds. It can be noted that both bounds provide an accurate estimation of the exact QCRB, while become tight rapidly as $N_s$ increases, which verifies our previous claim about the convergence of the bounds with a rate of \(\mathcal{O}(\eta N_s^{-1})\).

% We finally consider the fixed-budget scenario of
% Section~\ref{sec:photon_allocation}. We set
% \(N_{\rm tot}=500\), \(e=0\),
% \(\epsilon_{\rm SEP}=0.05\), and \(\tau_\Delta=0.04\) and $a,\beta=1/2$.  Fig.~\ref{fig:blocklength_delta} compares the numerical allocation
% limits with the closed-form results of
% Section~\ref{sec:photon_allocation}. We note that in the sequential scenario, \(n_\Delta\) known symbols are first used
% to estimate \(\Delta\), followed by \(n_{\rm c}\) communication
% symbols, with \(n=n_\Delta+n_{\rm c}\). The sensing limit specifies the
% minimum required acquisition length,
% \(n_\Delta\geq n_\Delta^{\rm suf}\), while the communication limit
% specifies the maximum total number of symbols supported by the fixed
% photon budget, \(n\leq n_{\rm SEP}^{\rm LG,int}\). If the two limits
% coincide, all allowable symbols are needed for sensing and none remain
% for communication. When
% \(n_{\rm SEP}^{\rm LG,int}>n_\Delta^{\rm suf}\), the shaded region is
% feasible and their difference gives the maximum communication
% allocation, \(n_{\rm c}^{\max}=n_{\rm SEP}^{\rm LG,int}-n_\Delta^{\rm suf}\).
% If the sensing limit exceeds the communication limit, no sequential
% allocation can satisfy both prescribed requirements.
We finally consider the scenario of Section~\ref{sec:photon_allocation} with \(N_{\rm tot}=500\), \(\epsilon_{\rm SEP}=0.05\), \(\tau_\Delta=0.04\), and \(\alpha=\beta=1/2\). In the sequential protocol, \(n_\Delta=\beta n\) known symbols are first used to estimate \(\Delta\), followed by \(n_{\rm c}=(1-\beta)n\) communication symbols. After accounting for the photon and symbol allocations, the sensing requirement imposes the lower limit \(n\geq n_\Delta^{\rm suf}\), whereas the SEP requirement imposes the upper limit \(n\leq n_{\rm SEP}^{\rm LG,int}\). The shaded region therefore contains, for a fixed pair\((\alpha,\beta)\), all the number of symbols \(n\) for which both requirements can hold. Each feasible value of \(n\) determines the corresponding sequential allocation \(n_\Delta=\beta n\) and \(n_{\rm c}=(1-\beta)n\). If the two limits coincide, only one $n$ is feasible, while \(n_\Delta^{\rm suf}>n_{\rm SEP}^{\rm LG,int}\) indicates that no sequential allocation can satisfy both targets. Importantly, the difference \(n_{\rm SEP}^{\rm LG,int}-n_\Delta^{\rm suf}\) represents the width of the feasible interval. A larger difference indicates greater flexibility in selecting the total number $n$, which implies that both thresholds are easier to jointly satisfy.

Figure~\ref{fig:blocklength_delta} compares the numerical limits with their closed-form counterparts. First, the sufficient sensing limit \(n_\Delta^{\rm suf}\) closely follows \(n_\Delta^{\rm num}\), showing that the moment-information
bound provides an almost exact, although slightly conservative,
prediction of the required number of acquisition symbols. Similarly,
\(n_{\rm SEP}^{\rm LG,int}\) follows \(n_{\rm SEP}^{\rm num}\) with a
relative error of approximately \(10\%\). This discrepancy arises
primarily from collapsing the two Gaussian tails in
\eqref{eq:sep_two_point_gl} into the effective tail in
\eqref{eq:sep_two_point_collapsed}. As the photon budget increases,
the minimum sensing length decreases and approaches a floor, whereas
the maximum number of symbols permitted by the SEP requirement
increases. The two limits therefore cross at the minimum photon budget
required for joint feasibility, after which the shaded feasible region
progressively widens and more symbols can be allocated to communication
following the estimation of \(\Delta\). The feasible region also
widens as \(\kappa\) increases because the von Mises phase distribution
becomes more concentrated. thus, degrading less
both sensing and communication performance.

\section{Conclusion}
In this work, we focused on the interplay between communication and phase estimation for a dephasing quantum channel with thermal noise. The derived SEP approximation and FI bounds provided tractable characterizations of this interplay and revealed the communication degradation due to inaccurate phase estimation,  and the sensing floor that stems from dephasing. Moreover, comparison of the heterodyne receiver's FI with the QFI showed that heterodyne detection can attain the quantum sensing limit as the number of photons increases. Based on these results, the conditions under which a communication and sensing requirement can jointly hold were given. Overall, our results revealed that increasing the photon budget suppresses the effects of thermal and measurement noise, but it cannot lift the fundamental limitations imposed by random dephasing. Therefore, reliable transmission depends heavily on the phase concentration and the accuracy of the receiver's phase estimation. We note that the present work is limited to BPSK transmission, known BPSK sensing symbols, and known dephasing concentration. Future work can alleviate any of these assumptions, consider the optimal resource allocation for sensing and communication, phase or higher order modulation, and alternative quantum receivers.

\appendices
\section{Proof of Proposition~\ref{prop:two_point_gl_sep}}
\label{app:two_point_gl_sep_proof}

Starting from the exact hyperplane-detector SEP and using
$\rho=\eta N_s/\sigma_{\mathrm{het}}^2$,
$e=\Delta-\vartheta$, and $u=\phi-\Delta$, we obtain
\begin{equation}
    P_{\rm e}
    =
    \frac{1}{2\pi I_0(\kappa)}
    \int_{-\pi}^{\pi}
    \e^{\kappa\cos u}
    Q\!\left(
        \sqrt{\rho}\cos(u+e)
    \right)du.
    \label{eq:app_sep_centered}
\end{equation}
Setting $x=u+e$ and using the $2\pi$-periodicity of the
integrand gives
\begin{equation}
    P_{\rm e}
    =
    \frac{1}{2\pi I_0(\kappa)}
    \int_{-\pi}^{\pi}
    \e^{\kappa\cos(x-e)}
    Q\!\left(
        \sqrt{\rho}\cos x
    \right)dx.
    \label{eq:app_sep_shifted}
\end{equation}
Let
$\mathcal P=(-\pi/2,\pi/2)$ and
$\mathcal N=(\pi/2,\pi]\cup[-\pi,-\pi/2)$, corresponding to
positive and negative detector projections, respectively. Mapping
$\mathcal N$ onto $\mathcal P$ by shifting the integration
variable by $\pm\pi$, and using
$\cos(x\pm\pi)=-\cos x$ and $Q(-z)=1-Q(z)$, yields
\begin{equation}
\begin{aligned}
    &\int_{\mathcal N}
    \e^{\kappa\cos(x-e)}
    Q\!\left(\sqrt{\rho}\cos x\right)dx
    \\
    &=
    \int_{\mathcal P}
    \e^{-\kappa\cos(x-e)}dx
    -
    \int_{\mathcal P}
    \e^{-\kappa\cos(x-e)}
    Q\!\left(\sqrt{\rho}\cos x\right)dx.
\end{aligned}
\label{eq:app_negative_region_split}
\end{equation}
The first term, including the normalization factor, is
\begin{equation}
    P_{\infty}
    =
    \frac{1}{2\pi I_0(\kappa)}
    \int_{-\pi/2}^{\pi/2}
    \e^{-\kappa\cos(x-e)}dx.
    \label{eq:app_phase_floor}
\end{equation}
Substituting \eqref{eq:app_negative_region_split} into
\eqref{eq:app_sep_shifted} and using
$\e^z-\e^{-z}=2\sinh z$ gives
\begin{equation}
\begin{aligned}
    P_{\rm e}-P_{\infty}
    & =\!
    \frac{1}{\pi I_0(\kappa)}
    \!\int_{-\pi/2}^{\pi/2}
    \!\sinh\!\left(
        \kappa\cos(x-e)
    \right)
    Q\!\left(
        \sqrt{\rho}\cos x
    \right)dx.
\end{aligned}
\label{eq:app_sep_sinh}
\end{equation}
Splitting the integral at zero, applying $x\mapsto-x$ to
the negative half, and using
\[
    \sinh(A+B)+\sinh(A-B)
    =
    2\sinh(A)\cosh(B),
\]
gives the following 
\begin{equation}
\begin{aligned}
    P_{\rm e}-P_{\infty}
    &=
    \frac{2}{\pi I_0(\kappa)}
    \int_0^{\pi/2}
    \sinh\!\left(
        \kappa\cos e\cos x
    \right)
    \\
    &\quad\times
    \cosh\!\left(
        \kappa\sin e\sin x
    \right)
    Q\!\left(
        \sqrt{\rho}\cos x
    \right)dx.
\end{aligned}
\label{eq:app_exact_paired_sep}
\end{equation}
To apply two-point GL quadrature, we map the set
$[0,\pi/2]$ onto the set $[-1,1]$ through the substitution
$x=\frac{\pi}{4}(\xi+1)$. The nodes
$\xi_{1,2}=\mp1/\sqrt{3}$, with weights
$w_1=w_2=1$, then, correspond to
\[
    x_1
    =
    \frac{\pi}{4}
    \left(
        1-\frac{1}{\sqrt{3}}
    \right),
    \qquad
    x_2
    =
    \frac{\pi}{4}
    \left(
        1+\frac{1}{\sqrt{3}}
    \right).
\]
Therefore it holds that,
\begin{equation}
\begin{aligned}
    P_{\rm e}-P_{\infty}
    &\approx
    \frac{1}{2I_0(\kappa)}
    \sum_{i=1}^{2}
    \sinh\!\left(
        \kappa\cos e\cos x_i
    \right)
    \\
    &\quad\times
    \cosh\!\left(
        \kappa\sin e\sin x_i
    \right)
    Q\!\left(
        \sqrt{\rho}\cos x_i
    \right).
\end{aligned}
\label{eq:app_two_point_result}
\end{equation}
Using $\kappa_{\rm c}=\kappa\cos e$, $\kappa_{\rm s}=\kappa\sin e$,
$c_i=\cos x_i$, and
\[
    \Gamma_i
    =
    \frac{
        \sinh(\kappa_{\rm c}\cos x_i)
        \cosh(\kappa_{\rm s}\sin x_i)
    }{
        2I_0(\kappa)
    },
\]
\eqref{eq:app_two_point_result} becomes $ P_{\rm e}(\rho)
    \approx
    P_{\infty}
    +
    \Gamma_1Q(c_1\sqrt{\rho})
    +
    \Gamma_2Q(c_2\sqrt{\rho})$,
which completes the proof.

\section{Proof of Proposition~\ref{prop:heterodyne_fim}}
\label{app:proof_heterodyne_fim}
In this appendix, we evaluate the Cartesian Fisher-information expectation in
\eqref{eq:het_fim_delta_cartesian}. Since the likelihood and score function
depend on the heterodyne receiver's observation through its magnitude and its angle relative
to \(\Delta\), we first use a polar coordinate transformation and normalize the
radial variable to obtain the dimensionless form in the proposition. Specifically, let us set
\[
    y_I=r\cos\beta,
    \qquad
    y_Q=r\sin\beta,
    \qquad
    d\bm y=r\,dr\,d\beta.
\]
Then,
\[
    y_I\cos\Delta+y_Q\sin\Delta
    =
    r\cos(\beta-\Delta),
\]
and
\[
    -y_I\sin\Delta+y_Q\cos\Delta
    =
    r\sin(\beta-\Delta).
\]
Substituting these identities, together with
\eqref{eq:het_likelihood_main} and \eqref{eq:score_delta_main}, into
\eqref{eq:het_fim_delta_cartesian}, and introducing
\[
    \lambda
    =
    \frac{\sqrt{\eta N_s}}{\sigma_{\rm het}^2}r,
    \qquad
    \rho
    =
    \frac{\eta N_s}{\sigma_{\rm het}^2},
\]
gives
\begin{equation}
\begin{aligned}
    J_{\Delta}^{\rm het}
    &=
    \frac{1}{2}
    \sum_{s\in\{-1,+1\}}
    \frac{\kappa^2\e^{-\rho/2}}
    {2\pi I_0(\kappa)\rho}
    \int_0^\infty
    \lambda^3\e^{-\lambda^2/(2\rho)}
    \\
    &\quad\times
    \int_{-\pi}^{\pi}
    I_0(R_s)q^2(R_s)
    \frac{\sin^2(\beta-\Delta)}{R_s^2}
    \,d\beta\,d\lambda,
\end{aligned}
\label{eq:app_fim_dimensionless}
\end{equation}
where
\[
    R_s
    =
    \sqrt{
        \kappa^2+\lambda^2+
        2s\kappa\lambda\cos(\beta-\Delta)
    }.
\]
We can see, that the two symbol-conditioned contributions are identical. Indeed, the
shift \(\beta-\Delta\mapsto\beta-\Delta+\pi\) maps the \(s=-1\)
integrand to the \(s=+1\) integrand. Hence, the symbol average may be
evaluated using only \(s=+1\). Finally, let us set \(\delta=\beta-\Delta\). Since the angular integrand is
\(2\pi\)-periodic, the interval remains \([-\pi,\pi)\), and
\[
    R=
    \sqrt{
        \kappa^2+\lambda^2+
        2\kappa\lambda\cos\delta
    }.
\]
Using    $I_0(R)q^2(R)
    =
    \frac{I_1^2(R)}{I_0(R)}$
in \eqref{eq:app_fim_dimensionless}, gives
\[
\begin{aligned}
    J_{\Delta}^{\rm het}
    &=
    \frac{\kappa^2\e^{-\rho/2}}
    {2\pi I_0(\kappa)\rho}
    \int_0^\infty
    \lambda^3\e^{-\lambda^2/(2\rho)}
    \left[
        \int_{-\pi}^{\pi}
        \frac{I_1^2(R)}{I_0(R)}
        \frac{\sin^2\delta}{R^2}
        \,d\delta
    \right]d\lambda,
\end{aligned}
\]
which gives \eqref{eq:jdd_dimless_main} and
\eqref{eq:A_delta_main} and the proof is complete.
\section{Proof of Proposition~\ref{prop:cpb}}
\label{app:cpb}
The likelihood of the heterodyne receiver's output can be factorized as
\begin{equation}
    p(\phi,\bm y\mid s;\Delta)
    =
    p_\Phi(\phi;\Delta,\kappa)
    p(\bm y\mid s,\phi).
    \label{eq:complete_phase_factorization}
\end{equation}
We note that the conditional likelihood $p(\bm y\mid s,\phi)$ does not
depend explicitly on $\Delta$. Therefore, if the latent phase $\Phi$
were to be directly observed, the likelihood of the heterodyne receiver would reduce to $p_\Phi(\phi;\Delta,\kappa)$. This defines
our complete-phase information benchmark. To proceed, let $U=\Phi-\Delta$. Using the von Mises pdf, the 
score function is given as
\begin{equation*}
    u_{\Phi,\Delta}=\kappa\sin U.
\end{equation*}
Using $\E[\cos(nU)]=A_n(\kappa)$,
$\E[\sin(nU)]=0$, and
$\sin^2U=[1-\cos(2U)]/2$, we obtain
\begin{equation}
    J_{\Phi,\Delta}
    =
    \kappa^2\E[\sin^2U]
    =
    \frac{\kappa^2}{2}[1-A_2(\kappa)],
    \label{eq:complete_phase_delta_entry}
\end{equation}
which gives \eqref{eq:Jphi_main}. Conditioned on the known transmitted symbol $s$, the observation model
forms the Markov chain   $  \Delta\longrightarrow\Phi\longrightarrow\bm Y$, where the mapping from $\Phi$ to $\bm Y$ is the parameter-independent
heterodyne channel. Therefore, by the data-processing inequality for
Fisher information,
$J_{\Delta}^{\rm het}\leq J_{\Phi,\Delta}$, which completes the proof.

\section{Proof of Proposition~\ref{prop:moment}}
\label{app:moment}

For a statistic \(T(\bm Y)\) with mean \(\bm\mu(\Delta)\),
covariance \(\mathbf C(\Delta)\), and mean derivative
\(\bm d_\Delta=\partial\bm\mu/\partial\Delta\), we can find a lower bound on the FI based on \cite{Stein}, as follows
\begin{equation}
    J_{\bm Y\mid s}(\Delta)
    \geq
    \bm d_\Delta^{\mathsf T}
    \mathbf C^{-1}
    \bm d_\Delta.
    \label{eq:moment_information_inequality}
\end{equation}
Let \(T(\bm Y)=\bm Y=[Y_I,Y_Q]^{\mathsf T}\). Conditioned on the
known symbol \(s\), its mean valuse is
\begin{equation}
    \bm\mu_{\bm Y\mid s}
    =
    s\sqrt{\eta N_s}A_1(\kappa)\bm e_\Delta.
    \label{eq:cartesian_mean_moment_bound}
\end{equation}
Next, let us define the following quantities
\begin{equation*}
\begin{aligned}
    C_{\rm r}
    &=
    \sigma_{\rm het}^2
    +
    \eta N_s
    \left[
        \frac{1+A_2(\kappa)}{2}
        -
        A_1^2(\kappa)
    \right],\\
    C_{\rm t}
    &=
    \sigma_{\rm het}^2
    +
    \frac{\eta N_s}{2}
    \left[
        1-A_2(\kappa)
    \right].
\end{aligned}
\end{equation*}
Using the trigonometric moments of the Von Mises distribution, the Cartesian covariance of the  statistic $T$ is given as
\begin{equation}
    \mathbf C
    =
    \begin{bmatrix}
        C_{\rm r}\cos^2\Delta+C_{\rm t}\sin^2\Delta
        &
        (C_{\rm r}-C_{\rm t})\sin\Delta\cos\Delta
        \\
        (C_{\rm r}-C_{\rm t})\sin\Delta\cos\Delta
        &
        C_{\rm r}\sin^2\Delta+C_{\rm t}\cos^2\Delta
    \end{bmatrix}.
    \label{eq:cartesian_covariance_moment_bound}
\end{equation}
These entries follow from
\(\operatorname{Var}(\cos U)
=[1+A_2(\kappa)]/2-A_1^2(\kappa)\),
\(\operatorname{Var}(\sin U)
=[1-A_2(\kappa)]/2\), and
\(\operatorname{Cov}(\cos U,\sin U)=0\), where
\(U=\Phi-\Delta\). The derivative of the mean is
\begin{equation}
    \bm d_\Delta
    =
    s\sqrt{\eta N_s}A_1(\kappa)
    \begin{bmatrix}
        -\sin\Delta\\
        \cos\Delta
    \end{bmatrix}.
    \label{eq:cartesian_delta_derivative}
\end{equation}
From \eqref{eq:cartesian_covariance_moment_bound},
\(\det(\mathbf C)=C_{\rm r}C_{\rm t}\). Direct substitution into
the inverse of the \(2\times2\) covariance matrix gives
\begin{equation}
    \bm d_\Delta^{\mathsf T}
    \mathbf C^{-1}
    \bm d_\Delta
    =
    \frac{\eta N_sA_1^2(\kappa)}{C_{\rm t}}
    =
    \frac{
        \rho A_1^2(\kappa)
    }{
        1+\frac{\rho}{2}[1-A_2(\kappa)]
    },
    \label{eq:moment_delta_quadratic_form}
\end{equation}
where $\rho =    \frac{\eta N_s}{\sigma_{\rm het}^2}$. Using \eqref{eq:moment_information_inequality} gives
\eqref{eq:Jmom_main}. The result is independent of the sign of \(s\);
hence, averaging the two symbol-conditioned inequalities gives
\eqref{eq:moment_delta_order} and completes the proof.

\section{Proof of Prososition \ref{prop:qfi_upper_bound}}
\label{prop:qcrb_upper}
We first recall the extended-convexity property of the QFI \cite{AlipourRezakhani2015}. For any
parameter-dependent continuous decomposition
\begin{equation}
    \rho(\Delta)
    =
    \int w(x;\Delta)\tau_x(\Delta)\,dx,
    \label{eq:exact_form}
\end{equation}
the QFI satisfies
\begin{align}
    J_{\mathrm Q,\Delta}[\rho(\Delta)]
    &\le
    J_{\mathrm C,\Delta}[w(x;\Delta)] +
    \int
    w(x;\Delta)
    J_{\mathrm Q,\Delta}[\tau_x(\Delta)]\,dx,
    \label{eq:qfi_extended_convexity_general}
\end{align}
where
\begin{equation}
    J_{\mathrm C,\Delta}[w(x;\Delta)]
    =
    \int
    w(x;\Delta)
    \left[
        \partial_\Delta\log w(x;\Delta)
    \right]^2dx
    \label{eq:qfi_mixing_fi}
\end{equation}
is the classical Fisher information of the mixing density. Thus, the
QFI of a mixture is no greater than the sum of the information carried
by its mixing density and the average QFI of its component states. We note that \(\rho_\mathrm{out}\) is not in the exact form of \eqref{eq:exact_form}. To proceed in this direction,  for an arbitrary \(q\in\mathbb{R}\), we make the following change of variable
\begin{equation*}
    x=\phi-(1-q)\Delta
\end{equation*}
in the integral of \eqref{eq:qfi_received_state}. It follows that
\begin{equation*}
    \phi=x+(1-q)\Delta,
    \qquad
    \phi-\Delta=x-q\Delta.
\end{equation*}
Using the \(2\pi\)-periodicity of the phase distribution and the
conditional states, \eqref{eq:qfi_received_state} can therefore be
written as
\begin{equation}
    \rho_\mathrm{out}
    =
    \int_{-\pi}^{\pi}
    p_\Phi(x;q\Delta,\kappa)
    \rho_{\mathrm{out}}
    \bigl(s\mid x+(1-q)\Delta\bigr)\,dx.
    \label{eq:qfi_split_mixture}
\end{equation}
The introduction of the parameter \(q\) changes only the decomposition of the state, while the the
density operator on the left-hand side remains the same. Then, by applying \eqref{eq:qfi_extended_convexity_general} to
\eqref{eq:qfi_split_mixture} gives
\begin{align}
    J_{\mathrm Q,\Delta}
    &\le
    J_{\mathrm C,\Delta}
    \!\left[
        p_\Phi(x;q\Delta,\kappa)
    \right]
    \nonumber\\
    &\quad+
    \int_{-\pi}^{\pi}
    p_\Phi(x;q\Delta,\kappa)
    J_{\mathrm Q,\Delta}
    \!\left[
        \rho_{\mathrm{out}}
        \bigl(s\mid x+(1-q)\Delta\bigr)
    \right]dx.
    \label{eq:qfi_extended_convexity_applied}
\end{align}
The mean direction of the mixing density in
\eqref{eq:qfi_split_mixture} is \(q\Delta\). Its score with respect to
\(\Delta\) is then
\begin{equation*}
    \partial_\Delta
    \log p_\Phi(x;q\Delta,\kappa)
    =
    q\kappa\sin(x-q\Delta).
\end{equation*}
Consequently, the  FI is given as
\begin{equation*}
    J_{\mathrm C,\Delta}
    \!\left[
        p_\Phi(x;q\Delta,\kappa)
    \right]
    =
    q^2J_{\Phi,\Delta}.
    \label{eq:qfi_weight_contribution}
\end{equation*}

Moreover, the phase of each displaced thermal component in
\eqref{eq:qfi_split_mixture} is $
    \theta_x(\Delta)
    =
    x+(1-q)\Delta$, while $
    \frac{\partial\theta_x}{\partial\Delta}
    =
    1-q$. Therefore, the QFI transformation rule under reparameterization yields
\begin{equation}
    J_{\mathrm Q,\Delta}
    \!\left[
        \rho_{\mathrm{out}}
        \bigl(s\mid\theta_x(\Delta)\bigr)
    \right]
    =
    (1-q)^2J_{\mathrm{dth}}.
    \label{eq:qfi_component_contribution}
\end{equation}
The right-hand side is independent of \(x\). Because
\(p_\Phi(x;q\Delta,\kappa)\) integrates to one,
\eqref{eq:qfi_extended_convexity_applied} becomes
\begin{equation}
    J_{\mathrm Q,\Delta}
    \le
    q^2J_{\Phi,\Delta}
    +
    (1-q)^2J_{\mathrm{dth}}.
    \label{eq:qfi_split_bound}
\end{equation}

Next, since \eqref{eq:qfi_split_bound} holds for every \(q\), we select the
value that minimizes its right-hand side. Let us define
\begin{equation}
    g(q)
    =
    q^2J_{\Phi,\Delta}
    +
    (1-q)^2J_{\mathrm{dth}}.
\end{equation}
Its first and second derivatives are respectively 
$
    g'(q)
    =
    2qJ_{\Phi,\Delta}
    -
    2(1-q)J_{\mathrm{dth}},$ and $
    g''(q)
    =
    2\left(
        J_{\Phi,\Delta}+J_{\mathrm{dth}}
    \right)>0$. Thus, the unique minimizing value of \eqref{eq:qfi_split_bound} is
\begin{equation}
    q^\star
    =
    \frac{
        J_{\mathrm{dth}}
    }{
        J_{\Phi,\Delta}+J_{\mathrm{dth}}
    }.
    \label{eq:qfi_optimal_split}
\end{equation}
Substitution into \eqref{eq:qfi_split_bound} gives
\begin{equation}
    J_{\mathrm Q,\Delta}
    \le
    \frac{
        J_{\Phi,\Delta}J_{\mathrm{dth}}
    }{
        J_{\Phi,\Delta}+J_{\mathrm{dth}}
    }
    =
    \left[
        \frac{1}{J_{\Phi,\Delta}}
        +
        \frac{1}{J_{\mathrm{dth}}}
    \right]^{-1}.
    \label{eq:qfi_harmonic_bound}
\end{equation}
Finally, by substituting \eqref{eq:qfi_latent_phase} and \eqref{eq:qfi_displaced_thermal} into \eqref{eq:qfi_harmonic_bound} yields
\begin{equation*}
    J_{\mathrm Q,\Delta}
    \le
    \left[
        \frac{1}{\kappa A_1(\kappa)}
        +
        \frac{2N_a+1}{4\eta N_s}
    \right]^{-1},
\end{equation*}
which proves \eqref{eq:qfi_upper_bound}. By a simple inversion we then get
\eqref{eq:qcrb_extended_convexity}.

\section{Proof of Proposition \ref{prop:qfi_lower_bound}}
\label{app:qfi_lower_bound}

For start, let us fix a local operating point \(\Delta_0\) and consider the quadrature
\begin{equation*}
    P_{\Delta_0}
    =
    \frac{
        \e^{-j\Delta_0}a
        -
        \e^{j\Delta_0}a^\dagger
    }{
        j\sqrt{2}
    }.
    \label{eq:qfi_tangent_quadrature}
\end{equation*}
Conditioned on \(\Phi=\phi\), its mean is
\begin{equation*}
    \mathbb{E}[P_{\Delta_0}\mid\Phi=\phi]
    =
    s\sqrt{2\eta N_s}
    \sin(\phi-\Delta_0).
    \label{eq:qfi_conditional_quadrature_mean}
\end{equation*}
Since \(\Phi\sim\operatorname{VM}(\Delta,\kappa)\),
\begin{equation*}
    \mathbb{E}
    [\sin(\Phi-\Delta_0)]
    =
    A_1(\kappa)\sin(\Delta-\Delta_0),
\end{equation*}
and therefore,
\begin{equation}
    \langle P_{\Delta_0}\rangle
    =
    s\sqrt{2\eta N_s}\,
    A_1(\kappa)\sin(\Delta-\Delta_0).
    \label{eq:qfi_quadrature_mean}
\end{equation}
Now, let \(X\) denote the outcome obtained by measuring
\(P_{\Delta_0}\), with its pdf denoted as $p_X(X;\Delta,\Delta_0)$. Thus, its score function with respect to the unknown
parameter \(\Delta\), can be defined as
$    u_\Delta(X;\Delta_0)
    =
    \partial_\Delta
    \log p_X(X;\Delta,\Delta_0)$. Since
\(\mathbb{E}[u_\Delta(X;\Delta_0)]=0\)  has to hold, we obtain
\begin{align}
    \partial_\Delta\mathbb{E}[X]
    &=
    \mathbb{E}
    [Xu_\Delta(X;\Delta_0)]
    \nonumber\\
    &=
    \mathbb{E}
    \left[
        (X-\mathbb{E}[X])
        u_\Delta(X;\Delta_0)
    \right].
\end{align}
The Cauchy-Schwarz inequality then gives
\begin{equation}
    \left|
        \partial_\Delta\mathbb{E}[X]
    \right|^2
    \leq
    \operatorname{Var}(X)
    J_{P_{\Delta_0},\Delta},
    \label{eq:qfi_classical_moment_inequality}
\end{equation}
where \(J_{P_{\Delta_0},\Delta}\) is the classical Fisher information
about \(\Delta\) generated by measuring \(P_{\Delta_0}\). Since the
QFI is the maximum classical Fisher information over all quantum
measurements \cite{BraunsteinCaves1994}, we get that
\begin{equation}
    J_{\mathrm Q,\Delta}
    \geq
    J_{P_{\Delta_0},\Delta}
    \geq
    \frac{
        |\partial_\Delta
        \langle P_{\Delta_0}\rangle|^2
    }{
        \operatorname{Var}(P_{\Delta_0})
    }.
    \label{eq:qfi_measurement_moment_bound}
\end{equation}
It is straightforward that the right-hand side of
\eqref{eq:qfi_measurement_moment_bound} is maximized by choosing
\(\Delta_0=\Delta\). This is the quadrature tangent to
the mean received displacement. Under this  choice, we have that
\begin{equation}
    \left|
        \partial_\Delta
        \langle P_{\Delta_0}\rangle
    \right|^2\Big|_{\Delta_0=\Delta}
    =
    2\eta N_s A_1^2(\kappa).
    \label{eq:qfi_quadrature_slope}
\end{equation}
In addition, the  quadrature variance of a displaced thermal state is
\(N_a+1/2\). Then, the law of total variance gives
\begin{equation}
\begin{aligned}
    \operatorname{Var}(P_{\Delta_0})
    &=
    \mathbb{E}
    \left[
        \operatorname{Var}
        (P_{\Delta_0}\mid\Phi)
    \right]
    \nonumber+
    \operatorname{Var}
    \left[
        \mathbb{E}
        (P_{\Delta_0}\mid\Phi)
    \right]\Big|_{\Delta_0=\Delta} \\
    & = N_a+\frac{1}{2}
    +
    \eta N_s[1-A_2(\kappa)].
    \label{eq:qfi_total_variance}
\end{aligned}
\end{equation}
By substitution into \eqref{eq:qfi_measurement_moment_bound} we get
% \begin{equation}
% \begin{aligned}
%     J_{\mathrm Q,\Delta}
%     &\geq
%     \frac{
%         2\eta N_s A_1^2(\kappa)
%     }{
%         N_a+\frac{1}{2}
%         +
%         \eta N_s[1-A_2(\kappa)]
%     }\\
%     & = \frac{2N_a+1}{
%         4\eta N_s A_1^2(\kappa)
%     }
%     +
%     \frac{1}{\kappa A_1(\kappa)}.
%     \label{eq:qfi_moment_bound_intermediate}
%     \end{aligned}
% \end{equation}
\begin{align}
    J_{\mathrm Q,\Delta}
    &\geq
    \frac{
        2\eta N_s A_1^2(\kappa)
    }{
        N_a+\frac{1}{2}
        +
        \eta N_s[1-A_2(\kappa)]
    }
    \nonumber\\
    &=
    \left[
        \frac{2N_a+1}{
            4\eta N_s A_1^2(\kappa)
        }
        +
        \frac{1}{\kappa A_1(\kappa)}
    \right]^{-1},
    \label{eq:qfi_moment_bound_intermediate}
\end{align}
which proves \eqref{eq:qfi_lower_bound}, while a simple inversion proves
\eqref{eq:qcrb_quadrature_moment}.

\bibliographystyle{IEEEtran}
\bibliography{bib2}

@ARTICLE{Stein,
  author={Stein, Manuel and Mezghani, Amine and Nossek, Josef A.},
  journal={IEEE Signal Process. Lett.}, 
  title={{A Lower Bound for the Fisher Information Measure}}, 
  year={2014},
  volume={21},
  number={7},
  pages={796-799},
  doi={10.1109/LSP.2014.2316008}}

@article{BraunsteinCaves1994,
  title = {{Statistical distance and the geometry of quantum states}},
  author = {Braunstein, Samuel L. and Caves, Carlton M.},
  journal = {Phys. Rev. Lett.},
  volume = {72},
  issue = {22},
  pages = {3439--3443},
  numpages = {0},
  year = {1994},
  month = {May},
  publisher = {American Physical Society},
  doi = {10.1103/PhysRevLett.72.3439},
  url = {https://link.aps.org/doi/10.1103/PhysRevLett.72.3439}
}

@article{Pinel2013,
  title = {{Quantum parameter estimation using general single-mode Gaussian states}},
  author = {Pinel, O. and Jian, P. and Treps, N. and Fabre, C. and Braun, D.},
  journal = {Phys. Rev. A},
  volume = {88},
  issue = {4},
  pages = {040102(R)},
  numpages = {5},
  year = {2013},
  month = {Oct},
  publisher = {American Physical Society},
  doi = {10.1103/PhysRevA.88.040102},
  url = {https://link.aps.org/doi/10.1103/PhysRevA.88.040102}
}

@article{AlipourRezakhani2015,
  title = {{Extended convexity of quantum Fisher information in quantum metrology}},
  author = {Alipour, S. and Rezakhani, A. T.},
  journal = {Phys. Rev. A},
  volume = {91},
  issue = {4},
  pages = {042104},
  numpages = {7},
  year = {2015},
  month = {Apr},
  publisher = {American Physical Society},
  doi = {10.1103/PhysRevA.91.042104},
  url = {https://link.aps.org/doi/10.1103/PhysRevA.91.042104}
}

@ARTICLE{Zhang2022TimeDivisionISAC,
  author={{Zhang, Qixun and Sun, Hongzhuo and Gao, Xinye and Wang, Xinna and Feng, Zhiyong}},
  journal={IEEE J. Sel. Areas Commun.}, 
  title={Time-Division ISAC Enabled Connected Automated Vehicles Cooperation Algorithm Design and Performance Evaluation}, 
  year={2022},
  volume={40},
  number={7},
  pages={2206-2218},
  doi={10.1109/JSAC.2022.3155506}}

@article{Zhu2023TimeDivision,
  title={{Time-division based integrated sensing, communication, and computing in integrated satellite-terrestrial networks}},
  author={Zhu, Xiangming and Wang, Hua and Yang, Zhaohui and Pham, Quoc-Viet},
  journal={Digit. Signal Process.},
  volume={143},
  pages={104262},
  year={2023},
  publisher={Elsevier}
}

@article{liu2020quantum,
  title={{Quantum Fisher information matrix and multiparameter estimation}},
  author={Liu, Jing and Yuan, Haidong and Lu, Xiao-Ming and Wang, Xiaoguang},
  journal={J. Phys. A: Math. Theor.},
  volume={53},
  number={2},
  pages={023001},
  year={2020},
  publisher={IOP Publishing}
}

@ARTICLE{pereg2022communication,
  author={Pereg, Uzi},
  journal={IEEE Trans. Inf. Theory}, 
  title={{Communication Over Quantum Channels With Parameter Estimation}}, 
  year={2022},
  volume={68},
  number={1},
  pages={359-383},
  doi={10.1109/TIT.2021.3123221}}

@INPROCEEDINGS{wang2022joint,
  author={Wang, Shi-Yuan and Erdoğan, Tuna and Pereg, Uzi and Bloch, Matthieu R.},
  booktitle={2022 IEEE Information Theory Workshop (ITW)}, 
  title={{Joint Quantum Communication and Sensing}}, 
  year={2022},
  volume={},
  number={},
  pages={506-511},
  doi={10.1109/ITW54588.2022.9965810}}

@INPROCEEDINGS{munar2024joint,
  author={Munar-Vallespir, Pere and Nötzel, Janis},
  booktitle={2024 IEEE 10th World Forum on Internet of Things (WF-IoT)}, 
  title={{Joint communication and sensing over the lossy bosonic quantum channel}}, 
  year={2024},
  volume={},
  number={},
  pages={1-6},
  doi={10.1109/WF-IoT62078.2024.10811180}}

@article{erdogan2025joint,
  title={Joint communication and sensing with bipartite entanglement over bosonic channels},
  author={Erdo{\u{g}}an, Tuna and Wang, Shi-Yuan and Su, Shang-Jen and Bloch, Matthieu},
  journal={arXiv preprint arXiv:2501.15675},
  year={2025}
}

@article{liu2024quantum,
  title = {{Quantum integrated sensing and communication via entanglement}},
  author = {Liu, Yu-Chen and Cheng, Yuan-Bin and Pan, Xing-Bo and Sun, Ze-Zhou and Pan, Dong and Long, Gui-Lu},
  journal = {Phys. Rev. Appl.},
  volume = {22},
  issue = {3},
  pages = {034051},
  numpages = {17},
  year = {2024},
  month = {Sep},
  publisher = {American Physical Society},
  doi = {10.1103/PhysRevApplied.22.034051},
  url = {https://link.aps.org/doi/10.1103/PhysRevApplied.22.034051}
}

@article{nikoloska2025variational,
  title={{Variational Quantum Integrated Sensing and Communication}},
  author={Nikoloska, Ivana and Simeone, Osvaldo},
  journal={arXiv preprint arXiv:2511.16597},
  year={2025}
}

@ARTICLE{krikidis2026quantum,
  author={Krikidis, Ioannis},
  journal={IEEE Wireless Commun. Lett.}, 
  title={{Quantum Optical Integrated Sensing and Communication With Homodyne BPSK Detection}}, 
  year={2026},
  volume={15},
  number={},
  pages={1375-1379},
  doi={10.1109/LWC.2026.3654015}}

@article{aspachs2009phase,
  title = {{Phase estimation for thermal Gaussian states}},
  author = {Aspachs, M. and Calsamiglia, J. and Mu{\~n}oz-Tapia, R. and Bagan, E.},
  journal = {Phys. Rev. A},
  volume = {79},
  issue = {3},
  pages = {033834},
  numpages = {12},
  year = {2009},
  month = {Mar},
  publisher = {American Physical Society},
  doi = {10.1103/PhysRevA.79.033834},
  url = {https://link.aps.org/doi/10.1103/PhysRevA.79.033834}
}

@article{vidrighin2014joint,
  title={{Joint estimation of phase and phase diffusion for quantum metrology}},
  author={Vidrighin, Mihai D and Donati, Gaia and Genoni, Marco G and Jin, Xian-Min and Kolthammer, W Steven and Kim, MS and Datta, Animesh and Barbieri, Marco and Walmsley, Ian A},
  journal={Nat. Commun.},
  volume={5},
  number={1},
  pages={3532},
  year={2014},
  publisher={Nature Publishing Group UK London}
}

@article{huang2024exact,
  title = {{Exact Quantum Sensing Limits for Bosonic Dephasing Channels}},
  author = {Huang, Zixin and Lami, Ludovico and Wilde, Mark M.},
  journal = {PRX Quantum},
  volume = {5},
  issue = {2},
  pages = {020354},
  numpages = {21},
  year = {2024},
  month = {Jun},
  publisher = {American Physical Society},
  doi = {10.1103/PRXQuantum.5.020354},
  url = {https://link.aps.org/doi/10.1103/PRXQuantum.5.020354}
}

@article{genoni2011optical,
  title = {{Optical Phase Estimation in the Presence of Phase Diffusion}},
  author = {Genoni, Marco G. and Olivares, Stefano and Paris, Matteo G. A.},
  journal = {Phys. Rev. Lett.},
  volume = {106},
  issue = {15},
  pages = {153603},
  numpages = {4},
  year = {2011},
  month = {Apr},
  publisher = {American Physical Society},
  doi = {10.1103/PhysRevLett.106.153603},
  url = {https://link.aps.org/doi/10.1103/PhysRevLett.106.153603}
}

@article{gong2025quantum,
  title = {{Quantum-enhanced change detection and joint communication detection}},
  author = {Gong, Zihao and Guha, Saikat},
  journal = {Phys. Rev. A},
  volume = {112},
  issue = {3},
  pages = {032604},
  numpages = {8},
  year = {2025},
  month = {Sep},
  publisher = {American Physical Society},
  doi = {10.1103/fpc3-lmql},
  url = {https://link.aps.org/doi/10.1103/fpc3-lmql}
}

@article{zuo2011performance,
  title={{Performance of phase compensated coherent free space optical communications through non-Kolmogorov turbulence}},
  author={Zuo, Linzi and Dang, Anhong and Ren, Yongxiong and Guo, Hong},
  journal={Opt. Commun.},
  volume={284},
  number={6},
  pages={1491--1495},
  year={2011},
  publisher={Elsevier}
}

@ARTICLE{zhang2021overview,
  author={Zhang, J. Andrew and Liu, Fan and Masouros, Christos and Heath, Robert W. and Feng, Zhiyong and Zheng, Le and Petropulu, Athina},
  journal={IEEE J. Sel. Topics Signal Process.}, 
  title={{An Overview of Signal Processing Techniques for Joint Communication and Radar Sensing}}, 
  year={2021},
  volume={15},
  number={6},
  pages={1295-1315},
  doi={10.1109/JSTSP.2021.3113120}}

@ARTICLE{liu2022integrated,
  author={Liu, Fan and Cui, Yuanhao and Masouros, Christos and Xu, Jie and Han, Tony Xiao and Eldar, Yonina C. and Buzzi, Stefano},
  journal={IEEE J. Sel. Areas Commun.}, 
  title={{Integrated Sensing and Communications: Toward Dual-Functional Wireless Networks for 6G and Beyond}}, 
  year={2022},
  volume={40},
  number={6},
  pages={1728-1767},
  doi={10.1109/JSAC.2022.3156632}}

@book{djordjevic2022quantum,
  author    = {Djordjevic, Ivan B.},
  title     = {Quantum Communication, Quantum Networks, and
               Quantum Sensing},
  publisher = {Academic Press},
  address   = {Cambridge, MA, USA},
  year      = {2022},
  isbn      = {978-0-12-822942-2}
}

@ARTICLE{hanzo2025quantum,
  author={Hanzo, Lajos and Babar, Zunaira and Cai, Zhenyu and Chandra, Daryus and Djordjevic, Ivan B. and Koczor, Balint and Xin Ng, Soon and Razavi, Mohsen and Simeone, Osvaldo},
  journal={Proc. IEEE}, 
  title={{Quantum Information Processing, Sensing, and Communications: Their Myths, Realities, and Futures}}, 
  year={2025},
  volume={113},
  number={9},
  pages={1024-1074},
  doi={10.1109/JPROC.2024.3510394}}

@ARTICLE{giani2025quantum,
  author={Giani, Andrea and Win, Moe Z. and Conti, Andrea},
  journal={IEEE J. Sel. Areas Inf. Theory}, 
  title={{Quantum Sensing and Communication via Non-Gaussian States}}, 
  year={2025},
  volume={6},
  number={},
  pages={18-33},
  doi={10.1109/JSAIT.2024.3491692}}

@book{gradshteyn2014table,
  title={{Table of integrals, series, and products}},
  author={Gradshteyn, Izrail Solomonovich and Ryzhik, Iosif Moiseevich},
  year={2014},
  publisher={Academic press}
}

@article{arekat2020fundamental,
  title={{The fundamental theorem of simple harmonic functions}},
  author={Arekat, Safwan},
  journal={Proc. Amer. Math. Soc. Ser. B},
  volume={7},
  number={8},
  pages={91--105},
  year={2020}
}

@article{FItutorial,
  title={{A tutorial on Fisher information}},
  author={Ly, Alexander and Marsman, Maarten and Verhagen, Josine and Grasman, Raoul PPP and Wagenmakers, Eric-Jan},
  journal={J. Math. Psychol.},
  volume={80},
  pages={40--55},
  year={2017},
  publisher={Elsevier}
}

@article{mele2024quantum,
  title = {{Quantum communication on the bosonic loss-dephasing channel}},
  author = {Mele, Francesco Anna and Salek, Farzin and Giovannetti, Vittorio and Lami, Ludovico},
  journal = {Phys. Rev. A},
  volume = {110},
  issue = {1},
  pages = {012460},
  numpages = {26},
  year = {2024},
  month = {Jul},
  publisher = {American Physical Society},
  doi = {10.1103/PhysRevA.110.012460},
  url = {https://link.aps.org/doi/10.1103/PhysRevA.110.012460}
}

@ARTICLE{monemi2026quantum,
  author={Monemi, Mehdi and Taskooh, Shiva Kazemi and Pourkarimi, Mohammad Reza and Rasti, Mehdi and Latva-aho, Matti and Hossain, Ekram},
  journal={IEEE Commun. Mag.}, 
  title={{Quantum Integrated Sensing and Communication: Foundations, Applications, and Future Directions}}, 
  year={2026},
  volume={},
  number={},
  pages={1-12},
  doi={10.1109/MCOM.001.2600359}}

@article{burenkov2021practical,
  title={{Practical quantum-enhanced receivers for classical communication}},
  author={Burenkov, IA and Jabir, MV and Polyakov, SV},
  journal={AVS Quantum Sci.},
  volume={3},
  number={2},
  year={2021},
  publisher={AIP Publishing}
}

@article{erkmen2009,
  title={{The classical capacity of single-mode free-space optical communication: A review}},
  author={Erkmen, Baris I and Moision, Bruce E and Birnbaum, Kevin M},
  journal={IPN Prog. Rep.},
  volume={42},
  pages={179},
  year={2009}
}

@article{fanizza2021squeezing,
  title={{Squeezing-enhanced communication without a phase reference}},
  author={Fanizza, Marco and Rosati, Matteo and Skotiniotis, Michalis and Calsamiglia, John and Giovannetti, Vittorio},
  journal={Quantum},
  volume={5},
  pages={608},
  year={2021},
  publisher={Verein zur F{\"o}rderung des Open Access Publizierens in den Quantenwissenschaften}
}

@INPROCEEDINGS{vonmises,
  author={Wang, Sichun and Read, William},
  booktitle={2025 IEEE Canadian Conference on Electrical and Computer Engineering (CCECE)}, 
  title={{On the von Mises Phase Model for the FFT-Based Two-Channel Digital Interferometer}}, 
  year={2025},
  volume={},
  number={},
  pages={449-453},
  doi={10.1109/CCECE64018.2025.11364389}}
\end{document}